# Has Science Established that the Cosmos is Physically Comprehensible?

Nicholas Maxwell
University College London
nicholas.maxwell@ucl.ac.uk

**Abstract**
   Most scientists would hold that science has not established that the cosmos is physically comprehensible – i.e. such that there is some as-yet undiscovered true physical theory of everything that is *unified*.  This is an empirically untestable, or metaphysical thesis.  It thus lies beyond the scope of science.  Only when physics has formulated a testable unified theory of everything which has been amply corroborated empirically will science be in a position to declare that it has established that the cosmos is physically comprehensible.
   But this argument presupposes a widely accepted but untenable conception of science which I shall call *standard empiricism*.  According to standard empiricism, in science theories are accepted solely on the basis of evidence.  Choice of theory may be influenced for a time by considerations of simplicity, unity, or explanatory capacity, but not in such a way that the universe itself is permanently assumed to be simple, unified or physically comprehensible.  In science, no thesis about the universe can be accepted permanently as a part of scientific knowledge independently of evidence.  Granted this view, it is clear that science cannot have established that the universe is physically comprehensible.
   Standard empiricism is, however, as I have indicated, untenable.  Any fundamental physical theory, in order to be accepted as a part of theoretical scientific knowledge, must satisfy two criteria.  It must be (1) sufficiently empirically successful, and (2) sufficiently *unified*.  Given any accepted theory of physics, endlessly many empirically more successful disunified rivals can always be concocted – disunified because they assert that *different* dynamical laws govern the diverse phenomena to which the theory applies.  These disunified rivals are not considered for a moment in physics, despite their greater empirical success.  This persistent rejection of empirically more successful but *disunified* rival theories means, I argue, that a big, highly problematic, implicit assumption is made by science about the cosmos, to the effect, at least, that the cosmos is such that all seriously *disunified* theories are false.
   Once this point is recognized, it becomes clear, I argue, that we need a new conception of science which makes explicit, and so criticizable and improvable the big, influential, and problematic assumption that is at present implicit in physics in the persistent preference for unified theories.  This conception of science, which I call *aim-oriented empiricism*, represents the assumption of physics in the form of a hierarchy of assumptions.  As one goes up the hierarchy, the assumptions become less and less substantial, and more and more nearly such that their truth is required for science, or the pursuit of knowledge, to be possible at all.  At each level, that assumption is accepted which (a) best accords with the next one up, and (b) has, associated with it the most empirically progressive research programme in physics, or holds out the greatest hope of leading to such an empirically progressive research programme.  In this way a framework of relatively insubstantial, unproblematic, fixed assumptions and associated methods is created, high up in the hierarchy, within which much more substantial and problematic assumptions and associated methods, low down in the hierarchy, can be changed, and indeed improved, as scientific knowledge improves.

One assumption in this hierarchy of assumptions, I argue, is that the cosmos is physically comprehensible – that is, such that some yet-to-be-discovered unified theory of everything is true.

Hence the conclusion: improve our ideas about the nature of science and it becomes apparent that science has already established that the cosmos is physically comprehensible – in so far as science can ever establish anything theoretical.

## 1. Introduction

Many scientists probably hold, as a matter of personal belief, that the cosmos is physically comprehensible. Most, however, would vehemently deny that science has already established that it is comprehensible. That will only be established when physics comes up with a "theory of everything" that is (a) unified, and (b) sufficiently empirically corroborated to be regarded as a part of theoretical scientific knowledge. The best candidate we have for such a theory is string theory (or M-theory, as it is sometimes called). But string theory has not yet received a satisfactory formulation, and it certainly has not been empirically corroborated. In the absence of a physical theory satisfying (a) and (b), all we have is the bald thesis that the cosmos is physically comprehensible – that is, the thesis that the cosmos is such that there is some yet-to-be-discovered theory of everything that is unified and true. But this thesis is metaphysical – too vague to be empirically testable.[1] Being metaphysical, it cannot be a part of theoretical scientific knowledge. It is incapable of being empirically corroborated, and it is actually incompatible with existing theoretical knowledge, in that our current best fundamental theories, the standard model and general relativity, are incompatible with one another, and thus incompatible with the assertion that there is some true, unified "theory of everything".

This argument seems decisive. But it is not. It presupposes a conception of science, called by me *standard empiricism*, which is untenable. Once the unacceptability of standard empiricism is appreciated, it becomes clear that we need to adopt a better conception of science, which I call *aim-oriented empiricism*. This holds that science has established that the cosmos is physically comprehensible – even though this thesis is metaphysical, and incompatible with current theoretical knowledge in physics.

The argument against standard empiricism (SE), and for aim-oriented empiricism (AOE), has been developed by me in a series of publications during the last forty years.[2] The argument began as an attempt to improve Karl Popper's falsificationist conception of science in order to overcome a difficulty fatal to that view. Subsequently it broadened into an argument which claimed to show that AOE is a dramatic improvement over all contending views about science. In one way, as we shall see, AOE is highly Popperian in character. But in other respects, it differs starkly from Popper's view, most notably in holding that metaphysical theses form an integral part of theoretical scientific knowledge.

Despite the many publications expounding the argument for AOE, the view itself, and the argument for it, have been almost entirely ignored so far by philosophers of science. Three criticisms of AOE have been published, two very bad,[3] one good,[4] but none valid.[5] Apart from that, books that I have published that expound and argue for AOE have received some glowing reviews.[6] Otherwise, silence.

Here, I will give as succinct a resumé as I can of the argument against SE and for AOE. I will then consider what the implications are of rejecting SE and accepting AOE in its stead for cosmology, for physics, for science more generally, for philosophy of science, and for the nature of rationality.

**2. Standard Empiricism: Exposition**

By "standard empiricism" I mean the doctrine that it is exclusively, or almost exclusively, evidence that decides what theories are accepted and rejected in science. Non-empirical considerations such as the simplicity, unity or explanatory character of a theory may influence choice of theory as well, for a time at least, but not in such a way that nature herself, or the phenomena, are presupposed to be simple, unified or comprehensible. *In science, no factual thesis about the world, or about the phenomena, can be accepted as a part scientific knowledge independently of empirical considerations*.

This rather thin doctrine is a central component of almost all views about science that philosophers of science have come up with. It is common ground for logical positivism, inductivism, logical empiricism, hypothetico-deductivism, conventionalism, constructive empiricism, pragmatism, realism, induction-to-the-best-explanationism, and the views of Karl Popper, Thomas Kuhn and Imre Lakatos.[7] There is a sense in which even Paul Feyerabend, and even social constructivist and relativist sociologists and historians of science uphold SE as the best available ideal of scientific rationality.[8] *If* science can be exhibited as rational, they hold (in effect), then this must be done in a way that is compatible with SE. The failure of science to live up to the rational ideal of SE is taken by them to demonstrate that science is not rational. That it is so taken demonstrates convincingly that they hold SE to be the only possible rational ideal for science (an ideal which cannot, it so happens, in their view, be met).

SE is more or less unthinkingly taken for granted by the vast majority of working scientists – so much so that it is rather rare to find the doctrine being explicitly formulated, let alone defended. Scattered throughout the writings of scientists one can, nevertheless, find affirmations of the view. Thus Planck once remarked "Experiments are the *only* means of knowledge at our disposal. The rest is poetry, imagination" (Atkins, 1983, p. xiv). Or, as Poincaré (1952, p. 140) put it "Experiment is the sole source of truth. It alone can teach us something new; it alone can give us certainty."[9]

Furthermore, scientists do what they can to ensure that science conforms to the doctrine. As a result, it exercises a widespread influence over science itself. It influences such things as the way aims and priorities of research are discussed and chosen, criteria for publication of scientific results, criteria for acceptance of results, the intellectual content of science, science education, the relationship between science and the public, science and other disciplines, even scientific careers, awards and prizes.[10]

**3. Standard Empiricism: Refutation**

If SE is valid, then the argument with which we began, which presupposes SE, is sound: it cannot be held that science has, today, established that the cosmos is physically comprehensible. But SE is not valid, as the following simple argument demonstrates.

Physics only ever accepts theories that are (more or less) *unified*, even though endlessly many empirically more successful disunified rivals can always be concocted. This persistent failure even to consider, let alone accept, empirically more successful disunified rivals means that physics makes a big, permanent assumption about the nature of the cosmos independent of all empirical considerations: the cosmos is such that all seriously disunified theories are false, whatever their empirical success might be. There is some kind of underlying unity in nature – to some extent at least. This in turn implies that SE is false, since SE asserts that science makes no persistent assumption about the cosmos independent of empirical considerations.

But is it the case that, given any accepted, unified theory, endlessly many empirically more successful, disunified rivals can always be concocted? Here is my demonstration that this is indeed the case.

Let T be any accepted fundamental physical theory – Newtonian theory, classical electrodynamics, quantum theory, general relativity, quantum electrodynamics, quantum electroweak theory, quantum chromodynamics, or the standard model. There are, to begin with, infinitely many disunified rivals to T – $T_1$, $T_2$, … $T_\infty$ – that are *just as empirically successful* as T. In order to concoct such a rival, $T_1$ say, all we need to do is modify T in an entirely *ad hoc* way for phenomena that occur after some future date. Thus, if T is Newtonian theory (NT), $NT_1$ might assert: everything occurs as NT predicts until the first moment of 2050 (GMT) when an inverse cube law of gravitation comes into operation: $F = Gm_1m_2/d^3$. Infinitely may such disunified rivals can be concocted by choosing infinitely many different future times for an abrupt, arbitrary change of law. These theories will no doubt be refuted as each date falls due, but infinitely many will remain unrefuted.[11] We can also concoct endlessly many disunified rivals to T by modifying the predictions of T for just one kind of system that we have never observed. Thus, if T is, as before, NT, then $NT_2$ might assert: everything occurs as NT predicts except for any system of pure gold spheres, each of mass greater than 1,000 tons, moving in a vacuum, centres no more than 1,000 miles apart, when Newton's law becomes $F = Gm_1m_2/d^4$. Yet again, we may concoct further endlessly many equally empirically successful disunified rivals to T by taking any standard experiment that corroborates T and modifying it in some trivial, irrelevant fashion – painting the apparatus purple, for example, or sprinkling diamond dust in a circle around the apparatus. We then modify T in an *ad hoc* way so that the modified theory, $T_3$ say, agrees with T for all phenomena except for the trivially modified experiment. For this experiment, not yet performed, $T_3$ predicts – whatever we choose. We may choose endlessly many different outcomes, thus creating endlessly many different modifications of T associated with this one trivially modified experiment. On top of that, we can, of course, trivially modify endlessly many further experiments, each of which generates endlessly many further disunified rivals to T.

Each of these equally empirically successful, disunified rivals to T – $T_1$, $T_2$, … $T_\infty$ – can now be modified further, so that each becomes *empirically more successful* than T. Any accepted fundamental physical theory is almost bound to face some empirical difficulties, and is thus, on the face of it, refuted – by phenomena A. There will be phenomena, B, which come within the scope of the theory but which cannot be predicted because the equations of the theory cannot (as yet) be solved. And there will be other phenomena (C) that fall outside the scope of the theory altogether. We can now take any one of the disunified rivals to T, $T_1$ say, and modify it further so that the new theory, $T_1^*$, differs further from T in predicting, in an entirely *ad hoc* way, that phenomena A, B and C occur in accordance with empirically established laws $L_A$, $L_B$ and $L_C$. $T_1^*$ successfully predicts all that T has successfully predicted; $T_1^*$ successfully predicts phenomena A that ostensibly refute T; and $T_1^*$ successfully predicts phenomena B and C that T fails to predict. On empirical grounds alone, $T_1^*$ is clearly more successful and better corroborated, than T.  And all this can be repeated as far as all the other disunified rivals of T are concerned, to generate infinitely many empirically more successful disunified rivals to T: $T_1^*, T_2^*, \ldots T_\infty^*$.[12]

But even though all of $T_1^*, T_2^*, \ldots T_\infty^*$ are more successful empirically than T, they are all, quite correctly, ignored by physics because they are all horribly disunified.  They postulate different laws for different phenomena in a wholly *ad hoc* fashion, and are just assumed to be false.  But this means physics makes a big, implicit assumption about the cosmos: it is such that all such grossly disunified, "patchwork quilt" theories are false.

If physicists only ever accepted theories that postulate atoms even though empirically more successful rival theories are available that postulate other entities such as fields, it would surely be quite clear: physicists implicitly assume that the cosmos is such that all theories that postulate entities other than atoms are false.  Just the same holds in connection with unified theories.  That physicists only ever accept unified theories even though endlessly many empirically more successful, disunified rival theories are available means that physics implicitly assumes that the cosmos is such that all such disunified theories are false.

But SE holds that no permanent thesis about the world can be accepted as a part of scientific knowledge independent of evidence (let alone against the evidence).  That physics does accept permanently (if implicitly) that there is some kind of underlying unity in nature thus suffices to refute SE.  SE is, in short, untenable.[13]  Physics makes a big implicit assumption about the nature of the cosmos, upheld independently of empirical considerations – even, in a certain sense, in violation of such considerations: the cosmos possesses some kind of underlying dynamic unity, to the extent at least that it is such that all sufficiently disunified physical theories are false. This is such a secure tenet of scientific knowledge that empirically successful theories that clash with it are not even considered for acceptance.

This argument establishes that SE is untenable.  It demonstrates that it is a part of current theoretical knowledge in physics that the cosmos is physically comprehensible to the extent, at least, that all sufficiently disunified, "patchwork quilt" physical theories are false.  This cosmological thesis, though metaphysical and thus not directly corroborated empirically, is nevertheless a central tenet of current scientific knowledge, so much so, indeed, that all physical theories (of the kind considered above) that clash with it are rejected whatever their empirical success might be.

We have taken a giant step towards our goal, but we are not there yet.  I have shown:

(A) Science accepts, as a part of theoretical knowledge, that the cosmos is such that all sufficiently disunified physical theories are false.

But what I need to show is the rather more substantial thesis:

(B) Science accepts, as a part of theoretical knowledge, that the cosmos is such that there is a yet-to-be-discovered physical "theory of everything" that is (a) unified, and (b) true.

In order to get from (A) to (B), two crucial questions need to be answered.

(1) What exactly does it mean to say of a physical theory that it is "unified"? How are degrees of unity, or disunity, to be assessed?

(2) Granted that science accepts, at least, that the cosmos is such that all sufficiently disunified theories are false, what metaphysical thesis about the cosmos ought science to adopt?

In what follows I take these two questions in turn.

### 4. Degrees of Disunity

What does it mean to say of a physical theory that it is *unified*? How can one capture this notion of the unity of a theory when an apparently beautifully unified theory can always be reformulated so that it becomes horribly disunified, and *vice versa*, a horribly disunified theory can be reformulated to become unified?[14] How are *degrees* of unity to be specified? These problems of "unification", or of "simplicity" as they are sometimes called, are widely understood to be fundamental problems of the philosophy of science.[15]

It is worth noting in passing that Einstein fully recognized the importance of the problem of what the unity of a theory is, and yet did not know how to solve it. This is apparent from a passage of Einstein's "Autobiographical Notes" during which he discusses ways in which physical theories, quite generally, can be critically assessed (Einstein, 1949, pp. 20-25).

Einstein emphasizes that theories need to be critically assessed from *two* distinct points of view: from the standpoint of their empirical success, and from the standpoint of their "inner perfection", the "naturalness" or "logical simplicity" of the postulates. Einstein stresses that the theories that need to be considered are those "whose object is the *totality* of all physical appearances", that is, "theories of everything" in modern parlance. And Einstein acknowledges that "an exact formulation" of the second point of view (having to do with the "inner perfection" of theories) "meets with great difficulties" even though it "has played an important role in the selection of theories since time immemorial". Einstein continues:

> The problem here is not simply one of a kind of enumeration of the logically independent premises (if anything like this were at all unequivocally possible), but that of a kind of reciprocal weighing of incommensurable qualities. ... The following I reckon as also belonging to the "inner perfection" of a theory: We prize a theory more highly if, from the logical standpoint, it is not the result of an arbitrary choice among theories which, among themselves, are of equal value and analogously constructed (Einstein, 1949, p. 23).

And Einstein comments:

> I shall not attempt to excuse the meagre precision of these assertions ... by lack of sufficient printing space at my disposal, but confess herewith that I am not, without more ado, and perhaps not at all, capable of replacing these hints with more precise definitions. I believe, however, that a sharper formulation would be possible (Einstein, 1949, p. 23).

Since Einstein's day, many attempts have been made to solve the problem, none successful.[16] How is the problem to be solved – a problem which baffled even Einstein?

The key to the solution is to appreciate that standard attempts to solve the problem have been looking at entirely the wrong thing. They have been looking at the *theory itself*, its axiomatic structure, its number of postulates, its formulation, its characteristic derivations, the language in which it is formulated. (Even Einstein made this mistake.) But all this is wrong. What one needs to look at is not the theory itself, but rather what the theory says about the world, the *content* of the theory in other words. One needs to look, not at the theory, but at the world *as depicted by the theory*. At a stroke the worst aspect of the problem of what unity *is* vanishes. No longer does one face what may be called the *terminological* problem of unity – the problem, namely, that the extent to which a theory is unified appears to be highly dependent on the way it is *formulated*. Suppose we have a given theory T, which is formulated in N different ways, some formulations exhibiting T as beautifully unified, others as horribly complex and disunified, but all formulations being interpreted in precisely the same way, so as to make precisely the same assertion about the world. If unity has to do exclusively with *content*, then *all these diverse formulations of T, having the same content, have precisely the same degree of unity*. The variability of apparent unity with varying formulations of one and the same theory, T, (given some specific interpretation) – which poses such an insurmountable problem for traditional approaches to the problem (see Salmon, 1989; Maxwell, 1998, pp. 56-68) – poses no problem whatsoever for the thesis that unity has to do with *content*. Variability of formulation of a theory which leaves its content unaffected is wholly irrelevant: the unity of the theory is unaffected.

But now we have a new problem: How is the unity of the *content* of a theory to be assessed? What exactly does it mean to assert that a dynamical physical theory has a unified content?

What it means is that the theory has *the same* content throughout the range of possible phenomena to which the theory applies. Unity, in other words, means that there is just *one* content throughout the range of possible phenomena to which the theory applies. If the theory postulates *different* contents, *different* laws, for different ranges of possible phenomena, then the theory is *disunified*, and the more such *different* contents there are so the more *disunified* the theory is. Thus "unity" means "one", and "disunity" means "more than one", the disunity becoming worse and worse as the number of different contents goes up, from two to three to four, and so on. Not only does this enable us to distinguish between "unified" and "disunified" theories; it enables us to assign "degrees of unity" to theories, or to partially order theories with respect to their degree of unity.[17]

All this can be illustrated by considering the disunified "patchwork quilt" theories indicated in Section 3 above. Whereas an accepted, unified theory, T, asserts that the same dynamical laws (as far as their content is concerned) apply in ranges of phenomena A, B and C, a "patchwork quilt" rival, T*, asserts that distinct laws, $L_A$, $L_B$, and $L_C$, apply in A, B and C, T* being disunified as a result.

To take a specific example, Newton's theory of gravitation, $F = GM_1M_2/d^2$ is unified in that what it asserts is *the same* throughout all possible phenomena to which it applies.

Bodies attract each other in proportion to their masses, and inversely proportional to the square of the distance between them, and this remains the same, according to the theory, throughout all the phenomena to which the theory applies, whatever the masses of the bodies may be, whatever their constitution, shape, relative velocity, distance apart, place and time. One kind of disunified version of Newton's theory would assert that Newton's law becomes an inverse cube law after some definite time $t_o$, the theory asserting that $F = GM_1M_2/d^2$ for times $t \leq t_o$ and $F = GM_1M_2/d^3$ for times $t > t_o$. This theory is disunified because what it asserts is blatantly *not* the same throughout the range of possible phenomena to which it applies.

Note that special terminology could be introduced to make Newtonian theory look *disunified*, and the disunified version of Newtonian theory look *unified*. All we need do is interpret "$d^N$" to mean "$d^N$ if $t \leq t_o$ and $d^{N+1}$ if $t > t_o$". In terms of this (admittedly somewhat bizarre) terminology, the disunified theory has the form "$F = GM_1M_2/d^2$", and Newtonian theory has the disunified *form* "$F = GM_1M_2/d^2$ for times $t \leq t_o$ and $F = GM_1M_2/d$ for times $t > t_o$". But this mere *terminological* reversal of disunity does not affect the *content* of the two theories: the content of Newtonian theory remains unified, and the content of the disunified version (which looks unified) remains disunified. For unity, in other words, we require that the theory is *terminologically* invariant throughout the range of possible phenomena to which it applies when *terminology*, used to formulate the theory, is itself invariant throughout the range of possible phenomena (so that *terminological* invariance implies *content* invariance).

This account of unity needs to be refined further. In assessing the extent to which a theory is disunified we may need to consider *how* different, or *in what way* different, one from another, the different contents of a theory are. A theory that postulates different laws at different times and places is disunified in a much more serious way than a theory which postulates the same laws at all times and places, but also postulates that distinct kinds of physical particle exist, with different dynamical properties, such as charge or mass. This second theory still postulates *different* laws for different ranges of phenomena: laws of one kind for possible physical systems consisting of one kind of particle, and slightly different laws for possible physical systems consisting of another kind of particle. But this second kind of difference in content is much less serious than the first kind (which involves different laws at different times and places). In other words, the more *different*, one form another, the different contents of a theory are, so the more seriously *disunified* the theory is.

What this means is that there are different *kinds* of disunity, different *dimensions* of disunity, as one might say, some more serious than others, but all facets of the same basic idea. Eight different facets of disunity can be distinguished as follows.

Any dynamical physical theory can be regarded as specifying a space, S, of possible physical states to which the theory applies, a distinct physical state corresponding to each distinct point in S. (S might be a set of such spaces.) For unity, we require that the theory, T, asserts that *the same* dynamical laws apply throughout S, governing the evolution of the physical state immediately before and after the instant in question. If T postulates N distinct dynamical laws in N distinct regions of S, then T has disunity of degree N. For unity in each case we require that N = 1.

(1) T divides space-time up into N distinct regions, $R_1...R_N$, and asserts that the laws governing the evolution of phenomena are the same for all space-time regions within each R-region, but are different within different R-regions. Example: the *ad hoc* version of Newtonian theory (NT) indicated above: for that theory, N = 2, in a type (1) way.[18]

(2) T postulates that, for distinct ranges of physical variables (other than position and time), such as mass or relative velocity, in distinct regions, $R_1,...R_N$ of the space of all possible phenomena, distinct dynamical laws obtain. Example: T asserts that everything occurs as NT asserts, except for the case of any two solid gold spheres, each having a mass of between one and two thousand tons, moving in otherwise empty space up to a mile apart, in which case the spheres attract each other by means of an inverse cube law of gravitation. Here, N = 2 in a type (2) way.

(3) In addition to postulating non-unique physical entities (such as particles), or entities unique but not spatially restricted (such as fields), T postulates, in an arbitrary fashion, M distinct, unique, spatially localized objects, each with its own distinct, unique dynamic properties. In this case, T is disunified to degree N = M + 1, in a type (3) way. Example: T asserts that everything occurs as NT asserts, except there is one object in the universe, of mass 8 tons, such that, for any matter up to 8 miles from the centre of mass of this object, gravitation is a repulsive rather than attractive force. The object only interacts by means of gravitation. Here, N = 2, in a type (3) way.

(4) T postulates physical entities interacting by means of N distinct forces, different forces affecting different entities, and being specified by different force laws. (In this case one would require one force to be universal so that the universe does not fall into distinct parts that do not interact with one another.) Example: T postulates particles that interact by means of Newtonian gravitation; some of these also interact by means of an electrostatic force $F = Kq_1q_2/d^2$, this force being attractive if $q_1$ and $q_2$ are oppositely charged, otherwise being repulsive, the force being much stronger than gravitation. Here, N = 2 in a type (4) way.

(5) T postulates N different kinds of physical entity,[19] differing with respect to some dynamic property, such as value of mass or charge, but otherwise interacting by means of the same force. Example: T postulates particles that interact by means of Newtonian gravitation, there being three kinds of particles, of mass m, 2m and 3m. Here, N = 3 in a type (5) way.

(6) Consider a theory, T, that postulates N distinct kinds of entity (e.g. particles or fields), but these N entities can be regarded as arising because T exhibits some symmetry (in the way that the electric and magnetic fields of classical electromagnetism can be regarded as arising because of the symmetry of Lorentz invariance, or the eight gluons of chromodynamics can be regarded as arising as a result of the local gauge symmetry of SU(3)). If the symmetry group, G, is not a direct product of subgroups, we can declare that T is fully unified; if G is a direct product of subgroups, T lacks full unity; and if the N entities are such that they cannot be regarded as arising as a result of some symmetry of T, with some group structure G, then T is disunified.[20] Example: T postulates the classical electromagnetic field, composed of the electric and magnetic fields, obeying Maxwell's equations for the field in the vacuum. The symmetry of Lorentz invariance unifies these two fields (see below). Here, N = 1 (in a type (6) way).

(7) If (apparent) disunity of there being N distinct kinds of particle or distinct fields has emerged as a result of cosmic spontaneous symmetry-breaking events, there being manifest unity before these occurred, then the relevant theory, T, is unified. If current (apparent) disunity has not emerged from unity in this way, as a result of spontaneous symmetry-breaking, then the relevant theory, T, is disunified. Example: Weinberg's and Salam's electroweak theory, according to which at very high energies, such as those that existed soon after the big bang, the electroweak force has the form of two forces, one with three associated massless particles, two charged, $W^-$ and $W+$, and one neutral, $W^o$, and the other with one neutral massless particle, $V^o$. According to the theory, the two neutral particles, $W^o$ and $V^o$, are intermingled in two different ways, to form two new, neutral particles, the photon, $\gamma$, and another neutral massless particle, $Z^o$. As energy decreases, the $W+$, $W-$ and $Z^o$ particles acquire mass, due to the mechanism known as spontaneous symmetry-breaking (involving the hypothetical Higgs particle), while the photon, $\gamma$, retains its zero mass. This theory unifies the weak and electromagnetic forces as a result of exhibiting the symmetry of local gauge invariance; this unification is only partial, however, because the symmetry group is a direct product of two groups, $U(1)$ associated with $V^o$, and $SU(2)$ associated with $W^-$, $W+$ and $W^o$.[21]

(8) According to general relativity, Newton's force of gravitation is merely an aspect of the curvature of space-time. As a result of a change in our ideas about the nature of space-time, so that its geometric properties become dynamic, a physical force disappears, or becomes unified with space-time. This suggests the following requirement for unity: space-time on the one hand, and physical particles-and-forces on the other, must be unified into a single self-interacting entity, U. If T postulates space-time and physical "particles-and-forces" as two fundamentally distinct kinds of entity, then T is not unified in this respect. Example: One might imagine a version of string theory without strings, different vibrational modes (perhaps) of empty, compactified six-dimensional space giving rise to the appearance of particles and forces, even though in reality there is only 10 dimensional space-time. Or one might imagine that the quantization of space-time leads to the appearance of particles and forces as only apparently distinct from empty space-time. In either case, N = 1: there is just the one self-interacting entity, empty space-time.

For unity, in each case, we require N = 1. As we go from (1) to (5), the requirements for unity are intended to be cumulative: each presupposes that N = 1 for previous requirements. As far as (6) and (7) are concerned, if there are N distinct kinds of entity which are not unified by a symmetry, whether broken or not, then the degree of disunity is the same as that for (4) and (5), depending on whether there are N distinct forces, or one force but N distinct kinds of entity between which the force acts.[22]

(8) does not introduce a new kind of unity, but introduces, rather, a new, more severe way of counting different kinds of entity. (1) to (7) require, for unity, that there is one kind of self-interacting physical entity evolving in a distinct space-time, the way this entity evolves being specified, of course, by a consistent physical theory. According to (1) to (7), even though there are, in a sense, two kinds of entity, matter (or particles-and-forces) on the one hand, and space-time on the other, nevertheless N = 1. According to (8), this would yield N = 2. For N = 1 in an (8) type way we require space-time and matter to be just one basic entity (unified by means of a spontaneously broken symmetry, perhaps).

As we go from (1) to (8), then, requirements for unity become increasingly demanding, with (6) and (7) being at least as demanding as (4) and (5), as explained above.[23]

(1) to (8) may seem very different requirements for unity. In fact they all exemplify the same basic idea: disunity arises when *different* dynamical laws govern the evolution of physical states in different regions of the space, S, of all possible physical states. For example, if a theory postulates more than one force, or kind of particle, not unified by symmetry, then in different regions of S different force laws will operate. If different fields, or different kinds of particle, are unified by a symmetry, then a symmetry transformation may seem to change the relative strengths of the fields, or change one kind of particle into another kind, but will actually leave the dynamics unaffected. In this case, then, the same fields, or the same kind of particle are, in effect, present everywhere throughout S. If (8) is not satisfied, there is a region of S where only empty space exists, the laws being merely those which specify the nature of empty space or space-time. The eight distinct facets of unity, (1) to (8) arise, as I have said, because of the eight *different* ways in which content can vary from one region of S to another, some differences being *more* different than others. Some of the above requirements for unity are suggested by developments in 20$^{th}$ century physics. This is true in particular of (6) to (8). The important point, however, is that all these requirements for unity, (1) to (8), exemplify the same basic idea: a theory, in order to be unified, must assert that the same laws apply throughout the possible phenomena to which it applies. (1) to (8) in effect represent different, increasingly subtle ways in which a theory can fail to be unified in this sense, granted that N > 1 in each case.

One point demonstrated by (1) to (8) is that there is no sharp distinction between the very crudest kind of unity or non-*ad hocness*, specified by (1) or (2), and the most demanding kind of theoretical unity, indicated by (8). One might, perhaps, draw a line between (1) to (3) on the one hand, and (4) to (8) on the other hand, and declare (1) to (3) to be essential to physics, and (4) to (8) as being desirable but optional: but any such distinction is more or less arbitrary.[24]

Symmetry plays a role in almost all of the above eight facets of unity.

What this role is, already indicated above, can be illustrated further by means of the example of classical electrodynamics and the symmetry of special relativity: Lorentz invariance. One might be inclined to hold that the electromagnetic field is made up of two distinct entities: the electric and the magnetic fields. But, as a result of being Lorentz invariant, the way the electromagnetic field divides up into the electric and magnetic fields depends on choice of reference frame. And, according to special relativity, nothing of fundamental dynamical significance can depend on choice of (inertial) reference frame. Any specific way we might choose to divide the field into the electric and magnetic fields would be arbitrary; it would be equivalent to choosing, arbitrarily, one frame, one state of motion to constitute absolute rest, from the infinitely many equally available, and all equivalent, according to special relativity. This would violate special relativity. Thus special relativity (Lorentz invariance) requires that we regard the electromagnetic field as *one* entity with two aspects, and not as *two* distinct entities.

I now consider briefly four questions that may be asked in connection with this proposed solution to the problem of unity of physical theory.

*First*, what of "simplicity"? Is this the same as "unity", or something distinct? The "simplicity" of a theory can be interpreted as having to do, not with whether the *same* laws apply throughout the space of possible phenomena predicted by the theory in question, but rather with the *nature* of the laws, granted that they are the same. Some laws are simpler than others. In order to overcome the objection that simplicity is formulation dependent it is essential, as in the case of unity, to interpret "simplicity" as applying to the *content* of theories, and not to their *formulation*, their *axiomatic structure*, etc. For details, see Maxwell (1998, pp. 157-9). It is a great success of the account of theoretical unity given here that it succeeds in distinguishing sharply between these two aspects of the problem of what the explanatory character of a physical theory *is*, namely the *unity* aspect, and the *simplicity* aspect, and succeeds in solving both.

On the face of it, mere terminological unity can play no important heuristic or methodological role in physics at all because, given any *unified* theory, it can be made as simple or complex as we like by appropriate choice of terminology. But what is paradoxical about the role of unity in physics is that terminological unity does, in practice, seem to be highly significant heuristically and methodologically. How is this paradox to be resolved? The answer is that, if terminological unity and *content* unity – or, as we may call it, *physical* unity – are in one-to-one correspondence with one another, so that the degree and kind of unity of one echoes precisely that of the other then, in these circumstances, terminological and physical unity are of equal significance (the former being an accurate guide to the latter). One way of arranging matters so that terminological and physical unity mirror each other is to ensure that symmetries of a theory can be interpreted in both "passive" and "active" senses – so that corresponding to a passive symmetry operation that is no more than a change in the way a physical system is described, there is a corresponding active symmetry operation that changes the physical system in question itself (changes its spatial position, orientation, or uniform velocity, for example, as opposed to the spatial position, orientation or velocity of the coordinate system used to describe the system).[25]

*Second*, does the question of whether laws governing a range of phenomena *remain the same* throughout their range of application have an unambiguous answer, in view of Goodman's "grue" and "bleen" paradox (Goodman, 1954)? Adapting Goodman's notions slightly, an object is grue if it is green up to the last moment of the 21[st] century, and blue thereafter; an object is bleen if it is blue up to the last moment of the 21[st] century, and green thereafter. Are not grue and bleen just as good predicates as blue and green? If the colours of objects change dramatically at the end of the 21[st] century, so that blue objects become green, and green objects blue, can we not, with equal legitimacy, say that there is no change, objects continue to be grue and bleen? This much discussed paradox is, in my view, very largely a red herring. On the face of it, the distinction made above, between formulation and content, suffices to dismiss the paradox. The sentence "This object is grue" (S) may, as far as its written form is concerned, be invariant through the end of the 21[st] century, but what this sentence asserts, its content, is not invariant. To this, the reply may be made that the *content* of S may be regarded as being invariant. But this is not what is ordinarily meant by "invariant" or "remain the same": the above account of unity of theory appeals to the ordinary meaning of "invariant" or "remains the same", and not the perverse grue and bleen meaning. Two additional points. It should be noted that the Goodman paradox implicitly accepts the ordinary meaning of "remains the same" in employing the terminology of "grue" and "bleen", terminology which remains the same, in the ordinary sense, throughout the end of the 21st century. Second, that the content of grue and bleen is not invariant with respect to the passage of time – unlike the content of blue and green which is invariant – is demonstrated by the point that if objects really are grue and bleen, and a person is convinced of this, then he can tell, by looking at grass and sky, whether or not the 21st century has ended, whereas the same is not true with respect to green and blue. Grue and bleen implicitly refer to a specific time in a way in which green and blue do not.

*Third*, Goodman's point concerning the ambiguity of "remains the same" may seem to gain support from the mathematical notion of a function as a rule which takes one from one set of numbers to another. According to this notion, the two functions (1) $y = 3x$ for all $x$, and (2) $y = 3x$ for $x \leq 2$ and $y = 4x$ for $x > 2$, are equally good functions. Both functions "remain the same" as $x$ increases and passes through the value $x = 2$. Clearly, we need a narrower notion of function than this if we are to be able to distinguish between functional relationships which do, and which do not, "remain the same" as values of variables change. We need to appeal to what may be called "invariant functions", functions which specify some fixed set of mathematical operations to be performed on "$x$" (or its equivalent) to obtain "$y$" (or its equivalent). In the example just given, (1) is invariant, but (2) is not. (2) is made up of two truncated invariant functions, stuck together at $x = 2$. Functions that appear in theoretical physics are *analytic*; that is, they can be represented as a power series (Penrose, 2004, pp. 112-4). Analytic functions are repeatedly differentiable. Such functions have the remarkable property that from any small bit of the function, the whole function can be reconstructed uniquely, by a process called "analytic continuation". All analytic functions are thus invariant. The latter notion is however a wider one, and theoretical physics might, one day, need to employ this wider notion explicitly, if space and time turn out to be discontinuous, and analytic functions have to be abandoned at a fundamental level.[26]

*Fourth*, can it really be meaningful to say that what a theory T asserts *remains the same* as one moves through a range of possible phenomena to which the theory applies? In order for this to make sense, does one not need to specify some aspect, some physical property or feature *with respect to which* what T asserts remains the same? The answer here is straightforward. It can be put like this. T is unified if what it asserts remains the same (as we move through possible phenomena to which T applies) *with respect to what determines (perhaps probabilistically) how phenomena evolve in space and time*. The aspect or feature with respect to which what T asserts must not change (for unity) is *that aspect of things which determines how events evolve*. Thus classical electrodynamics is unified because what it asserts about the electromagnetic field which determines the way the field evolves in space and time remains the same as we move through the space, S, of all possible states of the field.

This concludes my discussion of what it means to say of a theory that it is simple and unified.

This account of theoretical unity not only explicates and makes precise what it means to assert of a dynamical physical theory, T, that it is unified. It solves problems associated with the notion of theoretical unity, reveals that there are at least eight different kinds of unity, all exemplifying the same underlying notion, and provides the means for assessing theories with respect to their degree of unity. Furthermore, it explicates what we ought to mean when we declare of a physical theory that it is *explanatory*, for we can hold that the more (a) unified a theory is, and the greater its (b) empirical content (or predictive power), so the more *explanatory* it is.

And the above account of theoretical unity does more. It explicates what we ought to mean when we declare of the cosmos that it is *physically comprehensible*. And it brings order to the otherwise chaotic infinite realm of possible physically comprehensible universes – whether partially or wholly comprehensible. Let T stand for a "theory of everything" – a physical theory, that is, which, together with relevant initial and boundary conditions, in principle (not in practice) predicts accurately all physical phenomena (indeed, all physically possible phenomena). The thesis that the cosmos is at least partially physically comprehensible – a doctrine we may call *physicalism* - can be interpreted to mean that there is some yet-to-be-discovered theory of everything, T, which is (a) true, and (b) unified to some degree at least. But, as we saw above, there are eight different kinds of unity, n = 1, 2,… 8, each kind having some degree of unity, N = 1, 2 … ∞. For perfect unity, perfect physical comprehensibility, we require that n = 8 and N = 1. What the above account of theoretical unity enables us to do, then, is to throw a two-dimensional grid over all possible *partially physically comprehensible* universes – the grid partially ordering universes with respect to kind and degree of unity or physical comprehensibility. We can interpret physicalism(n,N) to assert: the cosmos is such that the true theory of everything, T, is disunified in an (n) type way to extent N, with n = 1 … 8, and N = 1 … ∞, the cosmos being perfectly physically comprehensible when n = 8 and N = 1.

## 5 Metaphysical Assumptions of Physics

In section 3 we saw that standard empiricism (SE) is untenable. In persistently accepting only *unified* theories even though empirically more successful disunified rivals can always be concocted, physics in effect makes a substantial metaphysical assumption about the cosmos: it is such that all seriously disunified theories are false. At once the all-important question arises: Given that physics accepts, as a part of theoretical knowledge, *some* assumption concerning the unity, the physical comprehensibility, of the cosmos – *some* version of physicalism – what precisely ought this assumption to be? In the last section we have seen that, given that the cosmos is physically comprehensible to some extent at least, there is, before us, a vast range of possibilities, as n varies from 1 to 8 (or 4 to 8), and N varies from 1 to some number less than 100 (we may perhaps assume). And there are, no doubt, infinitely many possible different partially or wholly physically comprehensible universes for any fixed values of n and N. What considerations ought to guide our choice, and what ought our choice to be?

*This is a question that it is both profoundly important for physics to answer correctly, and at the same time almost impossible to answer correctly.* The answer that physics gives – whether implicitly or explicitly – will have a massive affect over both the theories that theoretical physics seeks to develop, and the theories that physics accepts. Progress in theoretical physics is almost bound to be affected dramatically by how good the choice of assumption is that physics adopts concerning the physical comprehensibility of the cosmos. At the same time, it is almost impossible for us, in our present state of knowledge and ignorance, to give the correct answer, for what is at issue is that of which we are most ignorant, the ultimate nature of the cosmos.

Most physicists and cosmologists, all too aware of the impossibility of specifying correctly today the ultimate nature of the cosmos, even in general terms, come to the conclusion that no attempt should be made to do so (unless it can be formulated as a testable theory). But this is not an option, as we saw in section 3. If we really did allow evidence alone to determine what theories we accept, we would persistently accept empirically successful but horribly disunified theories, which would sabotage all hope of progress in theoretical physics and cosmology. In quite correctly excluding empirically successful but grossly disunified theories from consideration, and thus making progress in physics possible, inevitably a big implicit assumption about the nature of the cosmos is made.

The choice before us is thus not whether we should adopt some assumption about the ultimate nature of the cosmos, or refrain from making any such assumption. Rather it is whether we make the assumption we adopt *explicit*, so that it can be critically assessed, so that alternatives can be developed and criticized in the hope of *improving* the assumption we adopt or, on the other hand, whether we leave the assumption we adopt *implicit*, and thus uncriticizable, an unacknowledged dogma. Implicit, unacknowledged assumptions are much more likely to retard scientific progress than assumptions explicitly acknowledged for the obvious reason that the former cannot be subjected to critical scrutiny, to sustained attempts to improve them.

A glance at the history of physics bears this out. In practice, at any stage in its development, theoretical physics makes far more substantial and restrictive metaphysical assumptions about the nature of the cosmos than that indicated in section 3 above. These assumptions are made, either explicitly, or implicitly as a consequence of theoretical and mathematical concepts employed or symmetry principles adopted. The failure of physics to subject these implicit metaphysical assumptions to sustained criticism and attempted improvement has, on occasions, delayed discovery and acceptance of good new theories.

An example is Newton's theory of gravitation. When Newton put forward the theory in his *Principia* in 1686, the reigning metaphysical idea among the small band of natural philosophers at the time was the corpuscular hypothesis. This holds that nature is made up of tiny corpuscles which interact only by contact. This doctrine functioned as a standard of acceptability, even intelligibility: no fundamental physical theory could claim to be acceptable, even intelligible, if it could not be interpreted in terms of the corpuscular hypothesis. The impossibility of interpreting Newton's theory of gravitation in terms of the corpuscular hypothesis initially led some of Newton's most eminent contemporaries to reject Newton's theory. Thus Huygens, in a letter to Leibniz, writes: "Concerning the Cause of the flux given by M. Newton, I am by no means satisfied [by it], nor by all the other Theories that he builds upon his Principle of Attraction, which seems to me absurd. . . I have often wondered how he could have given himself all the trouble of making such a number of investigations and difficult calculations that have no other foundation that this very principle" (Koyré, 1965, pp. 117-8). Newton in a sense agreed, as is indicated by his remark: "That gravity should be innate, inherent and essential to matter, so that one body may act upon another, at a distance through a vacuum, without the mediation of anything else. . . is to me so great an absurdity, that I believe no man who has in philosophical matters a competent faculty of thinking can ever fall into it" (Burtt, 1932, pp. 265-6). The impossibility of interpreting the law of gravitation in terms of the corpuscular hypothesis, in terms of action-by-contact, led

Newton to interpret the law instrumentalistically, as specifying the way bodies move without providing any kind of explanation for the motion, in terms of unobservable forces.

   A somewhat more explicitly critical attitude towards the corpuscular hypothesis might have led Huygens and his contemporaries to appreciate that this view, far from embodying an ideal of comprehensibility, is actually a rather arbitrary special case of something much more general and comprehensible.  For the corpuscular hypothesis, viewed exclusively from the standpoint of physics, without any implicit appeal being made to intuitive notions (arising, for example, from familiarity with billiard balls), can be seen to be a very arbitrary special case of something much more general.  The corpuscle is a rigid sphere, on the surface of which there is an infinitely repulsive force that is zero elsewhere, so that if corpuscles collide they instantly rebound.  This can be regarded as an arbitrary special case of the much more general, and more mathematically and physically tractable, idea of a point-particle surrounded by a rigid, spherically symmetrical field of force which varies smoothly with distance from the central point-particle, alternatively attractive and repulsive perhaps at different distances, tending to infinitely repulsive as the centre is approached.  Just such an idea was eventually proposed by Boscovich in 1758.[27]  Reject the corpuscular hypothesis and adopt, instead, Boscovich's much more general, and more acceptable metaphysical view, and Newton's theory ceases to be incomprehensible and unacceptable, and becomes the very model of comprehensibility.[28]

   Another example is provided by James Clerk Maxwell's theory of electrodynamics. Maxwell himself, and most of his contemporaries and immediate successors, sought to interpret the electromagnetic field in terms of a material substratum, the hypothetical aether, itself to be understood in Newtonian terms.  A tremendous amount of effort was put into trying to understand Maxwell's field equations in terms of the aether.  Faraday, who appreciated that one should take the electromagnetic field as a new kind of physical entity, and explain matter in terms of the field rather than try to explain the field in terms of a kind of hypothetical matter (the aether), was ignored.  The unrealistic character, and ultimate failure, of mechanical models of the electromagnetic field led many to hold that the real nature of the field must remain a mystery.  The most that one could hope for from Maxwell's equations, it seemed, was the successful prediction of observable phenomena associated with electromagnetism.  This instrumentalistic attitude remained even after the advent of Einstein's special theory of relativity in 1905, which might be interpreted as giving credence to the idea that it is the field that is fundamental.  Gradually, however, the view came to prevail that one should see the field as a new kind of physical entity, quite distinct from the corpuscle and point-particle.

   Once again, misguided effort to interpret the new theory – in this case Maxwellian electrodynamics – in terms of the current metaphysics – Newtonian or Boschovichean point-atomism – might have been avoided if the then current metaphysics had been subjected to critical scrutiny.  For then it might have been realized – as Faraday evidently did realize – that Boscovich's point-atom view is an arbitrary special case of something more general.  Point-atomism assumes that the force field associated with each point-particle is rigid.  The point-atom remains at the centre of its associated force field however rapidly it accelerates in any direction.  A more general view would postulate that changes in the force field travel at some finite velocity – for example, the velocity of

light – rather than at infinite velocity. But this more general view becomes horrendously complicated if each point-particle continues to have, associated with it, its own distinct field. Granted Boscovichean point-atomism, in order to specify the instantaneous state of any physical system, one needs only to specify the instantaneous positions and velocities of the point-atoms of which the system consists. The infinite rigidity of the force-field of each point-atom means that the instantaneous state of the field, associated with any point-atom, is determined by the general law which species how the field varies with distance, plus the instantaneous state of the relevant point-particle. Once infinite rigidity is relaxed, however, this can no longer be done. The instantaneous state of each field, associated with each point-particle, must be specified independently for each spatial point. All this becomes much simpler if all the distinct fields of the distinct point-particles are amalgamated to form just one field – an entity that acquires an existence independent of the point-particles that are its source. Thus is the Faraday/Einstein notion of the field born. Subject Boscovichean point-atomism to critical scrutiny, to generalization, in other words, and one arrives at the metaphysical idea of the field which enables one to see Maxwellian electrodynamics as an entirely comprehensible, acceptable theory, exemplifying as it does the notion of the field. The need to interpret the theory in terms of the aether disappears.[29]

Much the same happened, in my view, in connection with quantum theory, as I will recount below.

It may be thought that the lesson to be learned from these episodes of bad, old, implicit metaphysics obstructing the acceptance and correct interpretation of a new theory is that physics should abjure *all* metaphysics, and insist that empirical considerations *alone* can decide whether a new theory is to be accepted or rejected. We have seen, however, in section 3, that it is just this option that is not possible. Non-empirical considerations must, inevitably, decide what theories are accepted and rejected, in addition, of course, to empirical considerations. Whatever these non-empirical considerations are, inevitably there will be metaphysical principles inherent in them, whether implicit or explicit. A vital key to rigour, success and progress in theoretical physics is to make implicit metaphysical assumptions explicit, so that they can be critically assessed and, we may hope, improved.

But how is this to be done? As I have already in effect remarked, the question of what metaphysical assumption about the nature of the cosmos to make is both *profoundly important* and *almost impossible for physics to get right*. On the one hand it is in our interests to choose an assumption that is as insubstantial as possible, to increase our chances of our chosen assumption being *true*. (The less you say, the greater the chances, other things being equal, of what you say being true.) But the more insubstantial the assumption that we choose, so the less well it helps constrain what theories we accept, and the avenues we pursue in seeking to develop new and better theories. On the other hand, if we choose an assumption that is very substantial, theories we accept and seek to develop will indeed be much more tightly constrained. This is excellent if the assumption is true, but a disaster if it is false, since it means we will constantly be accepting, and trying to develop, the wrong kind of theory. Progress in physics will be stultified. We must, and must not, make our choice of metaphysical assumption as substantial, and so as restrictive, as possible. How can we do justice to these conflicting desiderata?

# 6 Aim-Oriented Empiricism

The solution is to adopt, and put into scientific practice, a doctrine that I have called *aim-oriented empiricism*. The key idea of aim-oriented empiricism (AOE) is that we need to represent the metaphysical assumptions of physics in the form of a hierarchy of assumptions, as one goes up the hierarchy the assumptions becoming less and less substantial, and more nearly such that their truth is required for science, or the pursuit of knowledge, to be possible at all. In this way we create a relatively unproblematic framework of assumptions, and associated methodological rules, high up in the hierarchy, within which much more substantial, problematic assumptions, and associated methodological rules, low down in the hierarchy, can be critically assessed and, we may hope, improved, in the light of the empirical success they lead to, and other considerations: see figure 1.

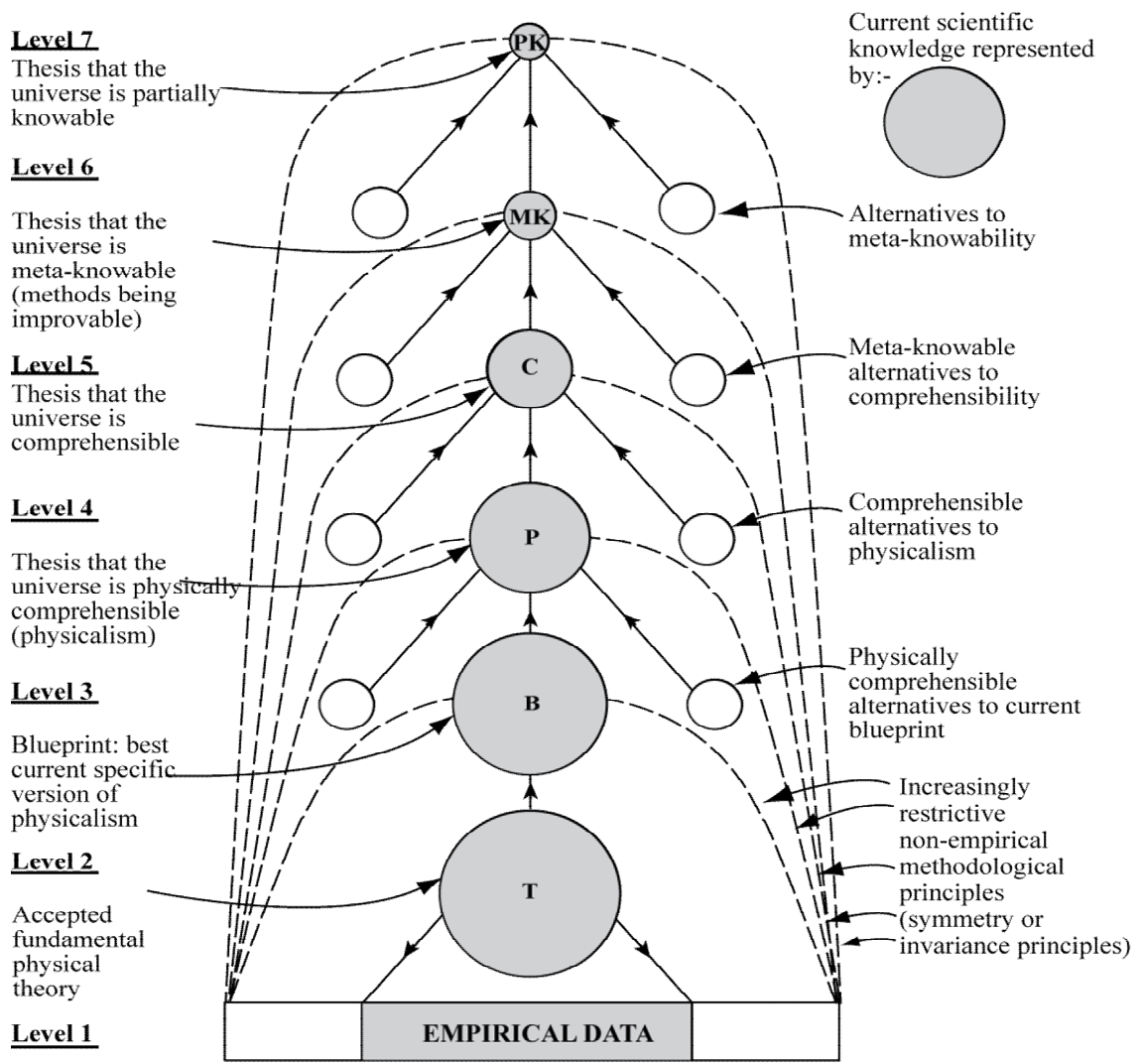

**Figure 1: Aim-Oriented Empiricism (AOE)**

At the top there is the relatively insubstantial assumption that the cosmos is such that we can acquire some knowledge of our local circumstances. If this assumption is false, we will not be able to acquire knowledge whatever we assume. We are justified in accepting this assumption permanently as a part of our knowledge, even though we have no grounds for holding it to be true. As we descend the hierarchy, the assumptions become increasingly substantial and thus increasingly likely to be false. At level 6 there is the more substantial thesis that there is some rationally discoverable thesis about the nature of the cosmos which, if true and if accepted, makes it possible progressively to improve methods for the improvement of knowledge. "Rationally discoverable", here, means at least that the thesis is not an arbitrary choice from infinitely many analogous theses. At level 5 we have the even more substantial thesis that the cosmos is *comprehensible* in some way or other, whether physically or in some other way. This thesis asserts that the cosmos is such that there is *something* (God, tribe of gods, cosmic goal, physical entity, cosmic programme or whatever), which exists everywhere in an unchanging form and which, in some sense, determines or is responsible for everything that changes (all change and diversity in the world in principle being explicable and understandable in terms of the underlying unchanging *something*). A cosmos of this type deserves to be called "comprehensible" because it is such that everything that occurs, all change and diversity, can in principle be explained and understood as being the outcome of the operations of the one underlying *something*, present throughout all phenomena. At level 4 we have the still more substantial thesis that the cosmos is *physically* comprehensible in some way or other. This asserts that the cosmos is made up one unified self-interacting physical entity (or one kind of entity), all change and diversity being in principle explicable in terms of this entity. What this amounts to is that the cosmos is such that some yet-to-be-discovered unified physical theory of everything is true. In terms of the terminology of section 4 above, this thesis is physicalism(8,1). At level 3, we have an even more substantial thesis, the best, currently available specific idea as to how the cosmos is physically comprehensible. This asserts that everything is made of some specific kind of physical entity: corpuscle, point-particle, classical field, quantum field, convoluted space-time, string, or whatever. Given the historical record of dramatically changing ideas at this level, and given the relatively highly specific and substantial character of successive assumptions made at this level, we can be reasonably confident that the best assumption available at any stage in the development of physics at this level will be false, and will need future revision. Here, ideas evolve with evolving knowledge. At level 2 there are the accepted fundamental theories of physics, currently general relativity and the standard model. Here, if anything, we can be even more confident that current theories are false, despite their immense empirical success. This confidence comes partly from the vast empirical content of these theories, and partly from the historical record. The greater the content of a proposition the more likely it is to be false; the fundamental theories of physics, general relativity and the standard model have such vast empirical content that this in itself almost guarantees falsity. And the historical record backs this up; Kepler's laws of planetary motion, and Galileo's laws of terrestrial motion are corrected by Newtonian theory, which is in turn corrected by special and general relativity; classical physics is corrected by quantum theory, in turn corrected by relativistic quantum theory, quantum field theory and the standard model.

Each new theory in physics reveals that predecessors are false. Indeed, if the level 4 assumption is correct, then all current physical theories are false, since this assumption asserts that the true physical theory of everything is unified, and the totality of current fundamental physical theory, general relativity plus the standard model, is notoriously disunified. Finally, at level 1 there are accepted empirical data, low level, corroborated, empirical laws.[30]

    The idea is to separate out what is most likely to be true, and not in need of revision, at and near the top of the hierarchy, from what is most likely to be false, and most in need of criticism and revision, near the bottom of the hierarchy. Evidence, at level 1, and assumptions high up in the hierarchy, are rather firmly accepted, as being most likely to be true (although still open to revision): this is then used to criticize, and to try to improve, theses at levels 2 and 3 (and perhaps 4), where falsity is most likely to be located.

    In order to be acceptable, an assumption at any level from 6 to 3 must (as far as possible) be compatible with, and a special case of, the assumption above in the hierarchy; at the same time it must be (or promise to be) empirically fruitful in the sense that successive accepted physical theories increasingly successfully accord with (or exemplify) the assumption. At level 2, those physical theories are accepted which are sufficiently (a) empirically successful and (b) in accord with the best available assumption at level 3 (or level 4). Corresponding to each assumption, at any level from 7 to 3, there is a methodological principle, represented by sloping dotted lines in the diagram, requiring that theses lower down in the hierarchy are compatible with the given assumption.

    When theoretical physics has completed its central task, and the true theory of everything, T, has been discovered, then T will (in principle) successfully predict all empirical phenomena at level 1, and will entail the assumption at level 3, which will in turn entail the assumption at level 4, and so on up the hierarchy. As it is, physics has not completed its task, T has not (yet) been discovered, and we are ignorant of the nature of the cosmos. This ignorance is reflected in clashes between theses at different levels of AOE. There are clashes between levels 1 and 2, 2 and 3, and 3 and 4. And the two fundamental theories at level 2, the standard model and general relativity, clash as well. The attempt to resolve these clashes drives physics forward.

    In seeking to resolve clashes between levels, influences can go in both directions. Thus, given a clash between levels 1 and 2, this may lead to the modification, or replacement of the relevant theory at level 2; but, on the other hand, it may lead to the discovery that the relevant experimental result is not correct for any of a number of possible reasons, and needs to be modified. In general, however, such a clash leads to the rejection of the level 2 theory rather than the level 1 experimental result; the latter are held onto more firmly than the former, in part because experimental results have vastly less empirical content than theories, in part because of our confidence in the results of observation and direct experimental manipulation (especially after repetition and expert critical examination). Again, given a clash between levels 2 and 3, this may lead to the rejection of the relevant level 2 theory (because it is disunified, *ad hoc*, at odds with the current metaphysics of physics); but, on the other hand, it may lead to the rejection of the level 3 assumption and the adoption, instead, of a new assumption (as has happened a number of times in the history of physics, as we saw in section 5 above). The rejection of

the current level 3 assumption is likely to take place if the level 2 theory, which clashes with it, is highly successful empirically, and furthermore has the effect of increasing unity in the totality of fundamental physical theory overall, so that clashes between levels 2 and 4 are decreased.  In general, however, clashes between levels 2 and 3 are resolved by the rejection or modification of theories at level 2 rather than the assumption at level 3, in part because of the vastly greater empirical content of level 2 theories, in part because of the empirical fruitfulness of the level 3 assumption (in the sense indicated above).

It is conceivable that the clash between level 2 theories and the level 4 assumption might lead to the revision of the latter rather than the former.  This happened when Galileo rejected the then current level 4 assumption of Aristotelianism, and replaced it with the idea that "the book of nature is written in the language of mathematics" (an early precursor of our current level 4 assumption).  The whole idea, however, is that as we go up the hierarchy of assumptions we are increasingly unlikely to encounter error, and the need for revision.  The higher up we go, the more firmly assumptions are upheld, the more resistance there is to modification.

The idea of representing the metaphysical presuppositions of physics (concerning the nature of the cosmos) as a hierarchy of theses, increasingly insubstantial as one goes up the hierarchy, gains some support from the fact that something somewhat similar exists informally at the empirical level – level 1 of the figure – and for much the same reason. There are, at the lowest level, the results of experiments performed at specific times and places. Then, above these, there are low-level experimental laws, asserting that each experimental result is a repeatable effect. Next up, there are empirical laws such as Hooke's law, Ohm's law or the gas laws. Above these there are such physical laws as those of electrostatics or of thermodynamics. And above these there are theories which have been refuted, but which can be "derived", when appropriate limits are taken, from accepted fundamental theory – as Newtonian theory can be "derived" from general relativity. This informal hierarchy at the empirical level exists for the same reason we need the hierarchy at the metaphysical level: so that relatively contentless and secure theses (at the bottom of the empirical hierarchy) may be distinguished from more contentful and insecure theses (further up the hierarchy) to facilitate pinpointing what needs to be revised, and how, should the need for revision arise. That such a hierarchy exists at the empirical level provides some support for my claim that we need to adopt such a hierarchy at the metaphysical level.

AOE, as depicted in figure 1, provides physics with a meta-methodology which facilitates improvement of the metaphysical assumptions, and associated methods, as physics advances, in the light of which seem to be the most fruitful empirically, and other considerations.  As knowledge in physics improves, so metaphysical assumptions and methods improve as well or, in other words, knowledge about how to improve knowledge.  There is something like positive feedback between improving knowledge of the cosmos, and improving aims and methods, improving knowledge about how to improve knowledge.  Everyone would acknowledge that this kind of positive feedback goes on at the empirical level.  New empirical knowledge can lead to new methods, via the development of new instruments, new experimental techniques, which in turn lead to further acquisition of new knowledge.  AOE provides methods which facilitate such positive feedback at the metaphysical and theoretical level as well.  As we increase our scientific knowledge and understanding of the cosmos, we increase our knowledge of how to increase knowledge – the very nub of scientific rationality which helps explain the explosive, and apparently ever accelerating, growth of scientific knowledge. We adapt the nature of science to what we find out about the nature of the cosmos. All this has gone on in science to some extent *implicitly*: what the transition from SE to AOE does is to make the implicit *explicit* – science becoming even more successful as a result.

A potential objection to what has been said so far is that, since Galileo and Kepler, physics has been extraordinarily successful in developing a succession of theories that satisfy the almost contradictory requirements of (a) enormously increasing the scope of empirical predictions, and simultaneously (b) bringing ever greater unity to fundamental theory in physics. This immense success has been achieved, for most of the time, with the physics community taking SE for granted. It is just not conceivable that this success could have been achieved with physicists accepting and implementing a conception of science that is irrational, unworkable and untenable, as section 3 above claims. The above refutation of SE must be invalid.

My reply, as I have just indicated, is that physics has met with such extraordinary success because AOE has been implicitly put into practice despite official allegiance being paid to SE. Success has been achieved *despite*, not *because of*, allegiance to SE. Fortunately, acceptance of SE has been somewhat hypocritical; it is that which has allowed something close to AOE to be put into scientific practice, which in turn has made it possible for physics to achieve such astonishing success. But, as I have indicated above, and as I shall point out again below, physics would be even more successful if it repudiated SE and explicitly adopted and implemented AOE. The attempt to keep alive the myth of SE obstructs, to some extent, the full implementation of AOE. It is of interest to note that Einstein, whose mature view of science included elements of AOE,[31] was aware of a mismatch between what theoretical physicists *say* they do, and what they *actually* do. He wrote "If you want to find out anything from the theoretical physicists about the methods they use … don't listen to their words, fix your attention on their deeds" (Einstein, 1973, p. 270).

As I have argued in some detail elsewhere (see note 31), AOE almost became explicit with Einstein's discovery of special and general relativity. Both theories arose out of Einstein's search for theoretical unity, the unification of Newton's and Maxwell's theories. In both cases Einstein was led to the new theory via the formulation of general principles: the restricted principle of relativity and the light postulate, in the case of special relativity, and Lorentz invariance and the principle of equivalence in the case of general relativity. Remarkably, Einstein first discovered general relativity in the form of a metaphysical idea: gravitation is the variably curvature of space-time induced by matter and energy. He then had to work hard to turn this into a precise, testable physical theory. This, according to AOE, is the manner in which new theories in physics should be created, as we shall see below. It is strikingly different from the way Newton and Maxwell created their theories: in these cases, not only did the then current metaphysical ideas not lead to the new theory – they actually obstructed the correct interpretation of the new theory once it had been formulated, as we have seen. Below I argue that this happened too with the discovery and interpretation of quantum theory. It is significant that Einstein believed passionately in the physical comprehensibility of the universe, and this belief played a significant role in his discovery of special and general relativity. He devoted his life to trying to discover how the cosmos is comprehensible, in the form of a testable, unified theory. And in his later years Einstein almost came to advocate a key component of AOE: physics cannot be understood without the assumption that the universe is physically comprehensible.[32] Eugene Wigner has pointed out, correctly, that the methods Einstein employed to discover his theories of relativity have had a profound impact on subsequent physics (Wigner, 1970, p. 15). But Einstein's work has not yet led to the repudiation of SE and the acceptance of AOE, as it perhaps should have done. It deserves to be noted, nevertheless, that the roots of AOE lie in physics, not in philosophy.[33]

**7 Post-Popperian Kantianism**

It has been established, in section 3 above, that physics must accept some metaphysical thesis concerning the (at least partial) physical comprehensibility of the cosmos. What arguments are there for holding that physics ought to accept, as a part of theoretical knowledge, that the cosmos is perfectly physically comprehensible as an integral part of AOE, as depicted in figure 1?

The first vital point to appreciate is that there is no argument to the effect that any of the metaphysical theses at levels 3 to 7 in figure 1 are *true*, or true with some degree of probability. They are all wholly conjectural in character. All that I attempt to establish is that physics is rationally entitled to accept these theses as a part of theoretical knowledge as the best available granted (a) physics seeks to improve knowledge of truth about the cosmos, and (b) some metaphysical thesis concerning the (at least partial) physical comprehensibility of the cosmos must be held to be a part of theoretical knowledge given that physics persistently only accepts unified theories, even though empirically more successful disunified rivals can always be concocted (and physics must accept some such metaphysical thesis in order to proceed at all).

I accept Karl Popper's point that, in the end, all our knowledge is conjectural in character – the vital task being to try to improve our conjectural knowledge by means of criticism.[34] There is, however, a crucial general consideration that Popper overlooked. We need to be *critically* critical and not, as Popper tended to be, merely *dogmatically* critical. The whole *raison d'etre* for criticism is to *improve* our knowledge, improve the extent to which it is truthful and explanatory. When criticism can be shown to be incapable of improving knowledge, it becomes irrational, and deserves to be ignored.

This has the important consequence that if we can show that there are items of conjectural knowledge which are such that, subjecting them to criticism cannot help the improvement of knowledge *in any circumstances whatsoever*, *even if they are false*, then there are rational grounds for holding that these items constitute permanent constituents of our knowledge *even though there is no argument or evidence to support their truth*. We have here the possibility of identifying what may be called *a priori conjectures* – items of conjectural knowledge which we are entitled to accept as a part of knowledge because we can show that accepting them as such can only aid, and can never obstruct or undermine, the search for knowledge *in any circumstances whatsoever*, *whether they are true or false*. It may be, for example, that a factual proposition is such that, if it is false, all factual knowledge becomes impossible. If so, then we have nothing to lose in accepting it as a part of knowledge even though we have no reason whatsoever to suppose it is true, because if it is false we cannot have any knowledge whatever we assume.

Other possibilities emerge from appreciating that the whole *raison d'etre* of criticism is to promote the growth, the improvement, of knowledge, criticism becoming irrational when it cannot do this. It may be that we can show that, *in present circumstances*, accepting some factual proposition as a part of our knowledge can only aid, and cannot hinder, the improvement of knowledge even if the proposition is false. We are rationally entitled to accept such a proposition as a part of our knowledge *as long as the relevant circumstances persist*. Again, it may be that we can show that subjecting items of knowledge P to criticism is more likely to contribute to the improvement of knowledge than is the criticism of items Q. In this case, we have rational grounds for directing our critical fire at P in preference to Q. Some items of conjectural knowledge become more conjectural than others. It may be that some thesis about the nature of the cosmos, though metaphysical and untestable, is nevertheless empirically fruitful in that it is associated with an empirically progressive research programme. Attempts to capture the thesis in the form of a testable theory have led to the development of a succession of theories that have met with great empirical success – and the thesis deserves to be accepted as a part of scientific knowledge on these grounds.

These are the kind of considerations I appeal to in arguing for AOE, and for the claim that science has already established that the cosmos is physically comprehensible.

The *critically* critical position I have just indicated might be called *post-Popperian Kantianism*. It is Popperian because it appreciates Popper's fundamental points that (a) all our knowledge is ultimately conjectural, and (b) we can nevertheless improve our conjectural knowledge by subjecting it to criticism. The position modifies Popperianism, however, in recognizing that some items of our conjectural knowledge play such a crucial role in the pursuit of knowledge that we are rationally entitled to accept them, and render them either entirely or partially immune to criticism, even though we have no grounds whatsoever for holding them to be true. These items of knowledge are quasi-Kantian in character, in that they amount to *a priori* items of knowledge, established by means of what might be called, after Kant, a "transcendental" argument, entirely, or almost entirely, independent of empirical considerations. On the other hand, these *a priori* items of knowledge are profoundly non-Kantian in character in that they are purely *conjectural* in character, and are not to any extent whatsoever known to be true (about our experience) with apodictic certainty – as Kant required of any "synthetic *a priori*" statement (a factual statement established by reason alone, by means of a transcendental argument). There is another crucial difference between Kant's and my *a priori* statements. Kant's were all about, not the real world (the *noumenal* world, as he called it) but the world of experience. The *a priori* statements with which we are concerned here are all conjectures about the *real* world – Kant's *noumenal* world.[35] And it is not just that the epistemological status and interpretation of the two sets of statements are quite different: the statements themselves are quite different – and the arguments for their acceptance are different.

There are, in short, elements of Kant and of Popper in *post-Popperian Kantianism*, but at the same time the doctrine differs dramatically from, and improves on, both Kant and Popper.

**8 Why Accept Aim-Oriented Empiricism?**
   I now spell out arguments for accepting the metaphysical theses of AOE, from level 7 to 4, beginning at the top. These arguments are all of the kind indicated in the previous section. They all stem from being *critically* critical. They seek to establish that we are rationally entitled to accept the thesis in question, as a part of theoretical scientific knowledge (even though we have no grounds for holding it to be true, or probably true), granted *some* metaphysical thesis is implicit in the methods of physics, and must be accepted. We are rationally entitled to accept the thesis in question because, for example, if it is false we can have no knowledge at all (and so acceptance can only help and cannot retard the pursuit of knowledge), or because it has led to a more empirically successful research programme than any rival thesis – or holds out a greater promise of doing so than any rival thesis.

**Level 7: Partial Knowability**. The cosmos is such that we can acquire some knowledge of our local circumstances. As I have already indicated, we are rationally entitled to accept this thesis because, if this thesis is false, we can acquire no knowledge whatsoever. Assuming the thesis to be true cannot in any way whatsoever, in any circumstances, sabotage or hinder the quest for knowledge. We are rationally entitled to accept the thesis as a permanent part of scientific knowledge even though we have no grounds for holding it to be true.

   It deserves to be noted that even our most trivial items of local, common sense knowledge make cosmological presuppositions. I would ordinarily be said to know that the desk now before me on which rests the computer I am using to write this sentence will continue to exist for the next two minutes. This very modest item of common sense knowledge about a particular object in my immediate surroundings nevertheless makes a substantial presupposition about the entire cosmos. It presupposes that, nowhere in the cosmos is an explosive event occurring which will spread with near infinite speed to engulf the earth, me, and my desk in 30 seconds time. In so far as we have any factual knowledge of anything, we have some knowledge of the entire cosmos. Nothing could demonstrate more dramatically Popper's point that *all* our knowledge is conjectural in character, however humble and restricted in scope it may be.

   Science undoubtedly corrects and refines common sense knowledge, but it also depends on it. Bereft of common sense knowledge of objects around us, we would not be able to set up and perform the sophisticated experiments and observations upon which modern physics and cosmology depend. Thus the cosmological presuppositions of ordinary common sense knowledge are also presuppositions of modern science.

   But might not science reveal that our local circumstances will soon become such that all knowledge, and life itself, will come to an end, perhaps because of the advent of a supernova in our immediate neighbourhood? Would not scientific knowledge then *negate* just that which it is supposed to presuppose? True. But our ability to acquire our knowledge of our imminent doom still depends on our local circumstances being such that we can continue to acquire knowledge of it. When our ability to acquire knowledge of our local circumstance disappears, we disappear too – and with us, science.

**Level 6: Meta-Knowability**. The cosmos is such that, not only can we improve our knowledge of it: we can improve our knowledge about how to improve knowledge. Or rather, slightly more modestly, the cosmos is such that, if we are deceived in our attempts to improve our knowledge about how to improve knowledge, we can always discover that we have been deceived. This does not guarantee that attempts to improve knowledge about how to improve knowledge will meet with success. But what it does guarantee is that if such attempts meet with empirical success, and the success is illusory, then we can discover, by empirical investigation, that it is illusory. As Einstein once put it "*Raffiniert ist der Herrgott, aber boshaft ist er nicht*", sometimes freely translated as "The universe is sublime, but not malicious".[36] The cosmos is not such that we are lured into thinking we are making great scientific progress when actually it is all illusory, and it would be quite impossible for us to discover that it is illusory (except, perhaps at some later date when the discovery may be catastrophic).

The idea, here, is that it may be within our power, given our existing knowledge, methods and capabilities, to formulate and adopt some "rationally discoverable" metaphysical thesis about the nature of cosmos which, if adopted as indicating the kind of theory physics should seek to develop and accept, provides the means to achieve genuine progress in theoretical knowledge in physics – the means to develop a genuinely empirically progressive research programme in physics. The idea, in other words, is that the cosmos may be knowable in the sense that knowledge about how to improve knowledge can be progressively improved, there being positive feedback between improving knowledge and improving knowledge about how to improve knowledge, along the lines promised by AOE – the final outcome being the true, physical theory of everything.

The notion of a cosmological thesis being "rationally discoverable" is problematic. It means, at least, that a "rationally discoverable" metaphysical thesis about the cosmos is one which can be formulated given our current knowledge and capabilities, and is such that it is not an arbitrary choice from infinitely many comparable, equally viable theses.

It is to be expected, of course that, as the hunt for the true theory of everything proceeds, AOE will be modified. Empirical data at level 1 will be amplified, new theories will emerge at level 2 which will expand on, correct and unify our current theories, the thesis at level 3 will, almost certainly, be modified, perhaps quite radically, and it is possible that this might happen to the thesis at level 4, and even, perhaps, at level 5. In order to take such future, and past, versions of AOE into account, we need to appeal to a view which I shall call *generalized aim-oriented empiricism* (GAOE). This asserts that we need to represent our knowledge and methods in physics in the form of a hierarchy, like the hierarchy of AOE, with the same theses as AOE at levels 7 and 6, and with the best available theses at levels 5, 4 and 3, the best available physical theories at level 2, and the best available empirical data at level 1.

We can now interpret meta-knowability to assert: the cosmos is such that there is some version of GAOE which we can formulate and implement which is such that, if it meets with empirical success and the success is illusory then, at any time, it is possible for us to discover that it is illusory.

I have two arguments in support of accepting meta-knowability as a part of scientific knowledge. First, however, I must consider an apparently lethal argument against AOE and GAOE, and thus against the basic thesis of this essay. Both AOE and GAOE hold that, as physics proceeds, we can progressively improve accepted metaphysical theses about the nature of the cosmos in the light of which seem best able to lead to the development of empirically successful theories, these theories being accepted in part because they accord with accepted metaphysical theses, there being positive feedback between improving metaphysical theses and empirically successful theories.

But this would seem to be a viciously circular procedure. We justify accepting a metaphysical thesis because it leads to an empirically successful theory, and then justify accepting the theory because it accords with our accepted metaphysical thesis. As Bas van Fraassen has put it, "From Gravesande's axiom of the uniformity of nature in 1717 to Russell's postulates of human knowledge in 1948, this has been a mug's game" (van Fraassen, 1985, pp. 259-60). How does AOE escape this charge?

One point can be made straight away. AOE makes no attempt to justify the truth (or probable truth) of any of its metaphysical theses. They are all irredeemably *conjectural* in character. A basic idea behind AOE is that metaphysical theses that are implicit in physics need to be made explicit so that they can be *criticized* – the very opposite of *justified*. A key feature of AOE is that theories, at level 2, are incompatible with metaphysical theses, at levels 3, 4 and 5. It hardly makes sense to claim the empirical success of theories justifies metaphysical theses when they are incompatible with them!

But even when all this has been conceded, there is still a problem. Even if AOE does not seek to justify the (probable) truth of any metaphysical thesis, it does seek to justify *acceptance* of these theses in the interests of the pursuit of truth.[37] And here we do seem to run into a version of the viciously circular argument which van Fraassen correctly condemns. Metaphysical thesis M is accepted because it supports acceptance of the empirically successful theory, T; and T is accepted, in part, because it is in accord with M. How can AOE be rescued from this apparently lethal charge of vicious circularity?

My proposed solution involves, crucially, an appeal to the level 6 thesis of meta-knowability.

Permitting metaphysical assumptions to influence what theories are accepted, and at the same time permitting the empirical success of theories to influence what metaphysical assumptions are accepted, may (if carried out properly), *in certain sorts of cosmos*, lead to genuine progress in knowledge. Meta-knowability is to be interpreted as asserting that *this is just such a cosmos*. And furthermore, quite essentially, reasons for accepting meta-knowability make no appeal to the success of science. In this way, meta-knowability legitimises the potentially invalid circularity of generalized AOE (GAOE), and of AOE.

Relative to an existing body of knowledge and methods for the acquisition of new knowledge, possible universes can be divided up, roughly, into three categories: (i) those which are such that the meta-methodology of GAOE or AOE can meet with no success, not even apparent success, in the sense that new metaphysical ideas and associated methods for the improvement of knowledge cannot be put into practice so that success (or at least apparent success) is achieved; (ii) those which are such that AOE can appear to be successful for a time, but this success is illusory, this being impossible to discover during the period of illusory success; and (iii) those which are such that GAOE, and even AOE, can meet with genuine success (it being possible to discover, at any time, that illusory success is indeed illusory, by ordinary methods of scientific investigation). Meta-knowability asserts that our universe is a type (i) or (iii) universe; it rules out universes of type (ii).

Meta-knowability asserts, in short, that the universe is such that AOE can meet with success and will not lead us astray in a way in which we cannot hope to discover by normal methods of scientific inquiry (as would be the case in a type (ii) universe). If we have good grounds for accepting meta-knowability as a part of scientific knowledge – grounds which do not appeal to the success of science – then we have good grounds for adopting and implementing generalized AOE (GAOE). Meta-knowability, if true, does not guarantee that GAOE will be successful. Instead it guarantees that GAOE will not meet with illusory success, the illusory character of this apparent success being such that it could not have been discovered by any means whatsoever before some date is reached.

If AOE lacks meta-knowability, its circular procedure, interpreted as one designed to procure knowledge to the extent that this is possible, becomes dramatically invalid, as the following consideration reveals. Corresponding to the succession of accepted fundamental physical theories developed from Newton down to today, there is a succession of severely disunified rivals which postulate that gravitation becomes a repulsive force from the beginning of 2150, let us say. Corresponding to these disunified theories there is a hierarchy of disunified versions of physicalism, all of which assert that there is an abrupt change in the laws of nature at 2150. The disunified theories, just as empirically successful as the theories we accept, render the disunified versions of physicalism just as scientifically fruitful as unified versions of physicalism are rendered by the unified theories we actually accept. The circularity inherent in AOE is invalid because it can be employed so as to lead to the adoption of *disunified* theories and metaphysical theses just as legitimately as it can be employed to lead to the adoption of *unified* theories and metaphysical theses. This is the case, at least, if AOE is bereft of meta-knowability. But if we have good reasons to accept meta-knowability as a part of scientific knowledge, then we have good reasons to reject *disunified* versions of physicalism: these lack the crucial requirement of rational discoverability. If we have good reasons to accept meta-knowability as an item of scientific knowledge, and these reasons make no appeal to the success of science, then the circularity inherent in AOE ceases to be invalid: meta-knowability asserts that the universe is such that empirical success achieved by implementing AOE will not be illusory *in a way which could not discovered by any means before a certain date*.

But what reasons have we for accepting meta-knowability that make no appeal to the success of science? One argument is simply this. If we accept meta-knowability, we have little to lose,[38] and may have much to gain. In accepting meta-knowability we decide, in effect, that it is worthwhile to try to improve knowledge about how to improve knowledge. We take seriously the possibility that the cosmos is such that we can discover something rather general about its nature which will enable us to *improve* our methods for improving knowledge. Not only do we hope to learn about the world; we hope to learn about how to learn about the world, and we are prepared to implement a meta-methodology (GAOE) which capitalizes on this possibility should it turn out to be actual. To fail to try to *improve* methods for improving knowledge on the grounds that apparent success might prove to be illusory is surely to proceed in a cripplingly over-cautious fashion. Any attempt at improving knowledge may unexpectedly fail, including the attempt to improve methods for improving knowledge. But eschewing the attempt to learn because it may fail cannot be sound: such an excuse for not making the attempt always exists. In accepting meta-knowability we do not assume, note, that the universe *is* such that GAOE will meet with success. We assume, merely, that it is such that *if* GAOE or AOE appears to meet with empirical success, this success will not be illusory in a way which could not have been discovered prior to the illusory character of the success becoming apparent. But this is an entirely sensible assumption to make. Nothing is to be gained from foregoing the attempt to acquire knowledge because of the fear that future, inherently unpredictable changes in the laws of nature may occur which render knowledge acquired obsolete.

   Can anything more be said? I think it can. We can argue as follows. As the pursuit of knowledge, and science, have developed over the millennia, GAOE has in fact been put into practice. Metaphysical presuppositions have been revised in the light of what seems to meet with the greatest empirical success – from myths, religious views, the ideas of the Presocratics, the ideas of Plato, Aristotle, Galileo, Boyle, Newton, and Boscovich, to metaphysical ideas concerning the underlying unity of the cosmos implicit in currently accepted symmetry principles, such as Lorentz invariance (associated with special relativity), gauge invariance (associated with quantum field theory and the standard model) and even, perhaps, supersymmetry (associated with string theory).  Even empirical methods have been revised in the light of metaphysical revisions. For example, given Aristotelian metaphysics, with its denial that precise mathematical laws govern natural phenomena, there is little point in performing precise experiments to decide between rival theories. This changes dramatically once Galileo's metaphysics is accepted, according to which "the book of nature is written in the language of mathematics" (an early statement of physicalism). Suddenly, it becomes highly pertinent to perform precise experiments, of the kind performed by Galileo involving, for example, rolling balls down inclined planes, to try to determine what precise mathematical law governs the fall of bodies near the surface of the earth. Granted, then, that GAOE has been put into practice over the millennia and right up to the present, science is more rigorous if the metaphysical assumption, implicit in this practice, is made explicit. This is the case even if this explicit thesis remains a conjecture *with no other reasons being given for its acceptance* over and above that it is implicit in scientific practice.  Meta-knowability is, in short, implicit in the scientific endeavour.  Science becomes more rigorous if this implicit assumption is made explicit, and is accepted as a part of scientific knowledge even though no other argument is forthcoming for its acceptance.[39]  Indeed, if this is not done, and meta-knowability is repudiated, then science becomes viciously circular, and thus inherently irrational – in so far as GAOE is implemented.  Instead of being directed against Gravesande and Russell, van Fraassen's "mug's game" argument becomes redirected, in an entirely valid way, against science itself!  In order to avoid this charge, science must accept meta-knowability as a (conjectural) item of knowledge, even if there is no other argument for its acceptance.[40]

   These two arguments, taken together, provide strong grounds for accepting meta-knowability as a permanent item of scientific knowledge.

**Level 5: Comprehensibility.** This is the thesis that the universe is comprehensible in some way or other, there being *something*, or an aspect of something (kind of physical entity, God, society of gods, cosmic purpose, cosmic programme or whatever) that runs through all phenomena, and in terms of which all phenomena can, in principle, be explained and understood. Almost all (perhaps all) cultures possess a myth, cosmology or religious view taken to explain natural phenomena, presupposed by attempts to improve knowledge. Almost all of these are personalistic, animistic or purposive in character: natural phenomena are explained in terms of the actions of gods, spirits, God, or purposes. Acceptance of some version of comprehensibility is often combined, however, with a clause that places strict limits on knowability (this clause being required, perhaps, to protect the thesis against criticism, and to explain away the lack of success of the view in promoting acquisition of knowledge). Thus God is said to be mysterious and unknowable. That the universe is held to be (more or less) comprehensible in almost all cultures is not, however, a good reason to hold it to be worthy of acceptance.

The comprehensibility thesis does, however, perfectly exemplify the basic idea of meta-knowability. Here, indeed, is a thesis about the nature of the cosmos which, if true, holds out the promise[41] that we can improve knowledge about how to improve knowledge by proposing and testing explanatory conjectures about phenomena, progressively modifying our idea as to how the cosmos is comprehensible in the direction of those explanatory conjectures which meet with the greatest empirical success.

Comprehensibility is, however, by no means the only thesis to exemplify the basic idea of meta-knowability. The cosmos might be meta-knowable even though only partially comprehensible, or even incomprehensible. It might be such that a series of theoretical revolutions are required in order to make progress, the number of distinct unified theories increasing by one each time there is a revolution, there being no end to this process. We might discover (or guess) that the cosmos has this structure after a few revolutions, and this guess might well help the endeavour to improve knowledge, even though the cosmos is ultimately incomprehensible.

What grounds are there for preferring the thesis of perfect comprehensibility to rival cosmological theses that also exemplify meta-knowability? Before the advent of modern science, they are very weak. We might argue that perfect comprehensibility is to be preferred because, if true, it promises to offer greater help with improving knowledge about how to improve knowledge than any rival comparable thesis. We might also argue that perfect comprehensibility seems intuitively more plausible than any partial comprehensibility thesis, or incomprehensibility thesis one can dream up. These other theses seem to come close to violating the requirement of "rational discoverability" of meta-knowability. To say that there are six, or twelve, distinct, basic entities that govern all phenomena – or six or twelve theories that can be unified no further – does seem to make an arbitrary choice between there being N such entities (or theories), N being any integer greater than one.

Before the advent of modern science, grounds for holding that the universe is perfectly comprehensible seem very weak. This is because no version of the conjecture led to spectacular empirically successful theorizing. It is only when modern science does exhibit spectacular empirically successful theorizing that strong grounds for holding that the cosmos is perfectly comprehensible do arise. It is only when we go down a level to *physical* comprehensibility that such strong grounds arise, as I now try to demonstrate.

**Level 4: Physicalism(8,1).** This is the thesis that the universe is physically comprehensible, everything being made up of just one kind of physical entity (or perhaps just one entity), all change and diversity being in principle explicable in terms of this one kind of entity. This thesis asserts that the universe is such that some yet-to-be-discovered physical "theory of everything" is unified in a type 8 way, and true.

Granted meta-knowability, we are justified in accepting that thesis, other things being equal, which holds out the greatest promise, if true, for progress in empirical knowledge. Physicalism(8,1) satisfies this requirement better than any rival thesis at this level, in that it places more demanding restrictions on any testable theory that is to be ultimately acceptable. (Such a theory must, in principle, predict and explain all physical phenomena, and must be unified in a type 8 way – the most demanding requirement for unity.) Physicalism(8,1) also indicates a path along which physics may proceed in order to improve empirical knowledge: testable theories need to be put forward and tested that, as far as possible (a) predict ever wider ranges of phenomena, and (b) are ever more unified. In order to develop good new theories, the attempt needs to be made to resolve clashes between existing empirically successful, unified theories. In short, physicalism(8,1), if true, indicates that AOE needs to be put into scientific practice.

But it is not just that physicalism(8,1) holds out the promise of progress; it has been associated, implicitly, with all the great advances in theoretical knowledge and understanding in physics at least since Galileo's time.

All advances in theory in physics since the scientific revolution have been advances in unification, in the sense of (1) to (8) above. Thus Newtonian theory (NT) unifies Galileo's laws of terrestrial motion and Kepler's laws of planetary motion (and much else besides): this is unification in senses (1) to (3). Maxwellian classical electrodynamics, (CEM), unifies electricity, magnetism and light (plus radio, infra red, ultra violet, X and gamma rays): this is unification in sense (4). Special relativity (SR) brings greater unity to CEM, in revealing that the way one divides up the electromagnetic field into the electric and magnetic fields depends on one's reference frame: this is unification in sense (6). SR is also a step towards unifying NT and CEM in that it transforms space and time so as to make CEM satisfy a basic principle fundamental to NT, namely the (restricted) principle of relativity. SR also brings about a unification of matter and energy, via the most famous equation of modern physics, $E = mc^2$, and partially unifies space and time into Minkowskian space-time. General relativity (GR) unifies space-time and gravitation, in that, according to GR, gravitation is no more than an effect of the curvature of space-time – a step towards unification in sense (8). Quantum theory (QM) and atomic theory unify a mass of phenomena having to do with the structure and properties of matter, and the way matter interacts with light: this is unification in senses (4) and (5). Quantum electrodynamics unifies QM, CEM and SR. Quantum electroweak theory unifies (partially) electromagnetism and the weak force: this is (partial) unification in sense (7). Quantum chromodynamics brings unity to hadron physics (via quarks) and brings unity to the eight kinds of gluons of the strong force: this is unification in sense (6). The standard model (SM) unifies to a considerable extent all known phenomena associated with fundamental particles and the forces between them (apart from gravitation): partial unification in senses (4) to (7). The theory unifies to some extent its two component quantum field theories in that both are locally gauge invariant (the symmetry group being U(1)xSU(2)xSU(3)). All the current programmes to unify SM and GR known to me, including string theory or M-theory, seek to unify in senses (4) to (8).[42]

In short, all advances in fundamental theory since Galileo have invariably brought greater unity to theoretical physics in one or other, or all, of senses (1) to (8): all successive theories have increasingly successfully exemplified and given precision to physicalism(8,1) to an extent which cannot be said of any rival metaphysical thesis, at that level of generality. The whole way theoretical physics has developed points towards physicalism(8,1), in other words, as the goal towards which physics has developed.

Furthermore, what it means to say this is given precision by the account of theoretical unity given in section 4 above.

In response to this claim it may be objected that theoretical physics could equally well be regarded as pointing towards a less restrictive version of physicalism – one which does not require matter and space-time to be unified, or one which demands only that the true theory everything is no more disunified than in a type 4 way to an extent N = 3, let us say (so that the true theory postulates three kinds of forces). What grounds are there for preferring physicalism(8,1) to physicalism(5,6), let us say? There are at least four, none of course decisive.

Fundamental to the whole argument for AOE is that physics needs to put into practice what may be called the *principle of intellectual integrity*: make explicit assumptions that are substantial, influential and implicit (AOE can be construed as the outcome of successive applications of this principle). In considering what thesis ought to be accepted at level 4, then, we need to consider what is implicit in those current methods of physics that influence what theories are to be accepted on non-empirical grounds – having to do with simplicity, unity, explanatoriness. There can be no doubt that, as far as non-empirical considerations are concerned, the more nearly a new fundamental physical theory satisfies all eight of the above requirements for unity, with N = 1, the more acceptable it will be deemed to be. Furthermore, failure of a theory to satisfy elements of these criteria is taken to be grounds for holding the theory to be false even in the absence of empirical difficulties. For example, high energy physics in the 1960s kept discovering more and more different hadrons, and was judged to be in a state of crisis as the number rose to over one hundred. Again, even though the standard model (the current quantum field theory of fundamental particles and forces) does not face serious empirical problems, it is nevertheless regarded by most physicists as unlikely to be correct just because of its serious lack of unity. In adopting such non-empirical criteria for acceptability, physicists thereby implicitly assume that the best conjecture as to where the truth lies is in the direction of physicalism(8,1). The principle of intellectual integrity requires that this implicit assumption – or conjecture – be made explicit so that it can be critically assessed and, we may hope, improved. Physics with physicalism(8,1) explicitly acknowledged as a part of conjectural knowledge is more rigorous than physics without this being acknowledged because physics pursued in the former way is able to subject non-empirical methods to critical appraisal as physicalism(8,1) is critically appraised, whereas physics pursued in the latter way cannot do this (or not to the same extent).

A second point to note is that it may well be that, even if some other version of physicalism(n,N) is true, with n < 8 and N > 1, nevertheless our best hope of discovering the truth may still lie in attempting to discover a theory that exemplifies physicalism(8,1), and failing in the attempt. As N becomes bigger, so the number of possible theories of everything compatible with that version of physicalism rapidly increases. (If N = 2, and the universe is made up of two distinct unified, dynamical patterns, there are, nevertheless, in general, infinitely many ways in which these two distinct patterns can be fitted together to make infinitely many different possible universes exemplifying just these two dynamic patterns. The step from one specified unified pattern to two is the step from one possible universe to infinitely many!) It makes sense to seek the simplest, most discoverable possibility, and design our methodology accordingly. As I have indicated, one can imagine a cosmos in which we might have reasons for adopting a methodological rule different from: (A) in order to be ultimately acceptable, a theory must be comprehensive and unified in a type (8) way. An example is: (B) in order to discover the true theory of everything, there need to be infinitely many theoretical revolutions, the number of forces increasing by one at each revolution. We cannot, therefore, just argue that, even if some version of physicalism other than physicalism(8,1) is true, nevertheless our best hope of discovering the truth is to adopt (A), try to discover a theory that exemplifies physicalism(8,1), and fail in the attempt. But we can argue that, in our current state of ignorance, our best bet is to adopt (A), and revise our acceptance of physicalism(8,1) if some other version of physicalism should emerge as appearing to fit the progress of physics better (such as a number of revolutions have taken place, and each time, the number of forces has gone up by one.)

There is another reason for preferring physicalism(8,1) to any other version, namely: only this version can do justice to the way general relativity unifies gravitation and space-time. This is a step towards type (8) unification in that, according to the theory, gravitation as a force disappears, and we are left with a dynamic theory of space-time. (Matter, or energy-density more generally, tells space-time how to curve: bodies then move along geodesics – the nearest things to straight lines in curved space-time.) It is above all general relativity which holds out the possibility that, not just gravitation, but all the forces and particles may be unified with space-time.

Finally, it needs to be remembered that what we are discussing is reasons for accepting physicalism(8,1) at level 4 within the context of AOE. If physicalism(8,1) was a candidate for the only metaphysical thesis to be accepted by science, it might well be thought to be much too specific and risky to be regarded as a part of scientific knowledge. But the whole point of AOE is that, as we descend the hierarchy, theses become increasingly specific, risky, tentative, and likely to require rejection, or at least revision. Physicalism(8,1) is bound to have a much more dubious epistemological status than partial knowability, let us say.

In section 3 we saw that physics must make some metaphysical assumption about the cosmos (whether this is explicitly acknowledged or not). In considering the acceptability of physicalism (8,1), then, what needs to be argued is not "why do we accept this thesis rather than none?", but rather "why is physicalism(8,1) more acceptable than its rivals?". One argument is that it accords better with meta-knowability than its rivals. A second argument is that it is implicit in current non-empirical requirements in physics for a theory to be unproblematically acceptable and – in accordance with the principle of intellectual integrity – this implicit assumption needs to be made explicit (so that it can be critically assessed and, if necessary, improved). A third argument is that the whole way theoretical physics has developed since Galileo down to today points towards physicalism(8,1). And a fourth argument is that, even if physicalism(8,1) is false, and n < 8 and N > 1, nevertheless it may well be that our best, even our only, hope of discovering the truth lies in assuming physicalism(8,1) and getting as close as we can to capturing this in a unified, empirically successful theory. If theoretical physics as a whole is only partially unified in that n < 8 and/or N > 1, the best – perhaps the only – grounds we can have for holding that nothing more unified exists is to strive, perhaps for centuries, to discover a more unified theory, and fail in the attempt. Only then would we have grounds for declaring, conjecturally, physicalism(8,1) to be false.

To sum up: some metaphysical thesis must be accepted as a part of theoretical scientific knowledge (as we saw in section 3). I have now shown that physicalism(8,1) is more acceptable than its rival metaphysical theses. I have shown that science has already established that the cosmos is physically comprehensible (to the extent that this is an item of scientific knowledge, and in so far as science can ever establish anything theoretical in character).[43]

I conclude this section with a few remarks about the above argument in support of this contention.

To begin with, I must re-emphasize two crucial features of what I claim to have established. First, there is no argument for the truth, or probable truth, of any of the theses at the various levels of AOE. All I have argued is that we are rationally entitled to accept these theses granted our aim in pursuing science is to improve knowledge about aspects of the cosmos, in so far as this is possible. In my view, we must accept Popper's point that all our knowledge is conjectural in character – even if some parts of our knowledge are more conjectural than others. Second, in section 3 we saw that physics, and therefore the whole of natural science, must accept some metaphysical thesis about the nature of the cosmos, whether this is acknowledged or not. What is at issue in this section, then, is not "what grounds are there for accepting any metaphysical thesis?", but rather "what grounds are there for preferring physicalism(8,1) to its rivals?". It is the second issue I have addressed in this section, not the first. Third, a remark about AOE. There are two inter-related reasons why we should adopt the hierarchical view of AOE, or GAOE (as opposed to accepting a metaphysical thesis or theses at one level as, for example, Bertrand Russell advocated). The epistemological standing of the various theses at the different levels of AOE are very different. Theses at levels 7 and 6 are such that it can never be in the interests of science, or the pursuit of knowledge, to repudiate them. The arguments in support of accepting these theses make no appeal to the empirical success of science. But theses at levels 5 and 4 might, conceivably be rejected in the future by science – especially the thesis at level 4. The arguments in support of accepting these theses do appeal to the empirical success of science, and to the whole character of this success. Only if the levels are distinguished can these very different reasons for accepting these different theses be distinguished and acknowledged. Furthermore, that the epistemological status of the theses at the different levels is different has everything to do with the heuristic and methodological fruitfulness of AOE (or GAOE). Not only does AOE concentrate criticism and the development of alternatives where it is likely to be most fruitful, at levels 3 and, to a lesser extent, 4. Furthermore, the epistemological security of theses at levels 7 and 6 – their acceptance being independent of the empirical success or failure of science (at levels 1 and 2) means that these theses provide permanent constraints on what can be accepted at lower levels. This is an aid to the improvement of theses at levels 3, and perhaps 4.

## 9. Implications

I now consider some implications of rejecting (all versions of) standard empiricism (SE), and accepting aim-oriented empiricism (AOE) instead.

(i) **Scope of Science**. There is a substantial increase in the scope of what we can legitimately take to *be* scientific knowledge. From the standpoint of standard empiricism, we do not (yet) possess scientific knowledge about the ultimate nature of the cosmos.

From the standpoint of aim-oriented empiricism, we do. As a result of moving from SE to AOE, the metaphysical thesis that the cosmos is physically comprehensible ceases to be a somewhat obscure, untestable, and therefore unscientific proposition, and becomes instead a lucid and secure item of theoretical scientific knowledge – more secure, indeed than any physical theory, such as quantum theory, however well corroborated empirically.

(ii) **Intellectual Significance**. We gain an item of scientific knowledge that is of profound significance, namely that the cosmos is physically comprehensible. From a general intellectual and cultural standpoint, many detailed, technical theoretical and observational discoveries of science are *not* of general significance – though of course some are (such as Darwin's theory of evolution). The scientific discovery (as we may choose to call it) that the cosmos is physically comprehensible is, like the theory of evolution, one of those discoveries that has implications for the whole of thought, and for all of our culture. It means, for example, that our whole human world, imbued with such things as colour, sounds, feelings, thoughts, consciousness, free will, meaning and value, must somehow be accommodated within the general framework of a physically comprehensible universe. Very general constraints are imposed on the way we can think about human life, its meaning, nature and value.[44]

(iii) **Quantum Theory**. As a result of rejecting SE and accepting AOE, the acceptability of orthodox quantum theory (OQT) is transformed. OQT is perhaps the most empirically successful theory ever formulated, given the great scope and variety of its predictions, and given that the theory remains unrefuted.[45] Granted SE, OQT must be judged to be acceptable. It satisfies in abundance the only scientifically relevant requirement – empirical success. But granted AOE, OQT must be judged to be unacceptable, or at least highly problematic, because of its disunity. As a result of the failure to solve the quantum wave/particle dilemma, OQT was developed as a theory which restricts itself to making predictions about the results of performing measurements on micro systems such as electrons and atoms. But this in turn means the theory that makes predictions is very seriously disunified in that it is made up of two mutually contradictory parts: a purely quantum mechanical part, and a part which describes the measuring instrument and consists, as Niels Bohr always emphasized,[46] of some part of classical physics.[47] OQT is, in other words, disunified in a type (2) way, as formulated in section 4 above. This is a very severe kind of disunity. It means, granted AOE, that OQT is at best highly problematic, even though so highly successful empirically.

This very serious defect in OQT stems, as Einstein always emphasized,[48] from the abandonment of micro-realism, the failure to solve the wave/particle problem, required if quantum theory is to be interpreted to be about micro systems per se, and how they interact with one another, whether they are being measured or not. This failure not only results in OQT being severely disunified; it also means, as I have argued in detail elsewhere, that OQT (a) fails to specify the physical nature of micro systems, (b) is ambiguous about whether nature is fundamentally probabilistic or deterministic, (c) lacks precision, (d) lacks explanatory power, (e) obstructs unification with general relativity, and (f) is incapable of being applied to the cosmos as a whole.[49]

I suggest the failure to develop an unproblematic version of QT, even though over 80 years elapsed since it was first formulated in 1925 and 1926, is due to the failure to put AOE into practice, and subject implicit metaphysical assumptions to explicit criticism.

The key question "is the electron a wave or particle?" implicitly presupposes determinism, since wave and particle are classical notions, and classical physics, at the fundamental level, is deterministic. But perhaps what QT is trying to tell us is that nature is fundamentally probabilistic. QT, after all, in general makes *probabilistic* predictions. If nature is fundamentally probabilistic, then the question "is the electron a wave or particle?" is entirely the wrong question to ask. The correct question is twofold. What kinds of unproblematic, fundamentally probabilistic entities are there, as possibilities? Can quantum entities – electrons and atoms – be construed to be some variety of unproblematic probabilistic entity – an entity which interacts with others in a probabilistic way? Elsewhere, I have demonstrated that this can be done.[50] All such fundamentally probabilistic versions of QT, however, face the problem of specifying, in precise, quantum mechanical terms, the conditions for probabilistic transitions to occur, no appeal being made to measurement, macroscopic phenomena or irreversibility. My specific proposal, here, is that probabilistic transitions occur when new particles or bound states are created as a result of inelastic interactions. Superpositions of different particle states exist, but do not persist. The outcome is a fully micro realistic, fundamentally probabilistic version of QT, free of the disunity defects of OQT, able to recover all the empirical success of OQT, and able to make as yet untested empirical predictions that differ from those of OQT.[51]

   My specific version of probabilistic QT may well not be correct. That the option of developing testable fundamentally probabilistic versions of QT has, by and large, been neglected is, however, in my view, somewhat disgraceful.[52] It is due, I suggest, once again to the failure to put AOE into explicit scientific practice, resulting in the failure to examine critically implicit metaphysical assumptions influencing interpretations of QT. Adopting a fundamentally probabilistic metaphysics transforms the interpretative problems of QT, but this requires that metaphysical ideas are subjected to imaginative and critical scrutiny within physics, something that SE discourages as "unscientific". In addition, of course, SE encourages the attitude that OQT, given its immense empirical success, cannot conceivably be regarded as inherently problematic when judged from a scientific standpoint. General acceptance of SE, and the failure to adopt and implement AOE, may well, in other words, have delayed the development of an unproblematic version of QT.

   The issue just discussed concerning QT has major implications, it should be noted, for subsequent developments in theoretical physics and cosmology – for quantum electrodynamics, for the standard model, for string theory, and for quantum cosmology.
(iv) **Implications for Cosmology**. Does AOE have more specific implications for cosmology? Might AOE help decide, for example, whether or not the cosmological theory that there are many, perhaps infinitely many, universes alongside ours is to be accepted as the true "theory of everything"? Might AOE have implications for the acceptability of string theory, or M-theory as it is sometimes called?

   The answer is, I think, yes. We might develop a potential "theory of everything", T, that meets with empirical success and is perfectly unified except that T predicts that the cosmos evolves into infinitely many universes, each having slightly different values of 30 or so parameters from any other, there being theoretical reasons why this cannot be interpreted as a cosmological probabilistic jump into just one of these infinitely many universes (so that none of the others exist). In this case we would be confronted by two theories, T and T*, both empirically adequate in our universe. The two theories differ in two respects. First, T asserts that the cosmos consists of infinitely many universes whereas T* asserts that there is just our universe. Second, whereas T is perfectly unified, T* is, in comparison, drastically disunified in that it has 30 constants (having to do with the strength of different forces and the properties of particles) which are entirely arbitrary, and have to be determined empirically. In these circumstances, we might take AOE to imply that T is to be accepted on the grounds that T* is drastically less unified than T. By contrast, SE implies that there is no scientific basis for preferring T to T* since both fit all empirical data equally well.

As far as string theory is concerned, we may take AOE to imply that, for perfect unity, we require there to be a unification of space-time and matter so that both are aspects of one underlying entity (possibly a kind of space-time such that matter is merely an aspect of it). Current formulations of string theory do not satisfy this requirement for perfect unity (so that physicalism(8,1) is exemplified). Strings are entities *in* space-time, they are not *aspects* of space-time. String theory does not, for example, postulate that space-time itself has a "stringy" aspect which gives rise to the strings that constitute particles of matter. Thus, according to AOE, quite apart from empirical considerations, string theory, as at present formulated, is inadequate because it fails to be fully unified – a version of physicalism(8,1).

(v) **Best Level 3 Blueprint**. AOE highlights the fundamental importance, for theoretical physics and cosmology, of formulating and adopting the best possible thesis – or "blueprint" – at level 3 in the hierarchy of theses of AOE, taking into account developments at levels 1, 2 and 4. One possibility is to accept the best available formulation of string theory. But this, as we have seen, contradicts the best available thesis at level 4, namely physicalism(8,1). Perhaps, in any case, a thesis somewhat less specific than string theory is required. One possibility is a doctrine that may be called *Lagrangianism*. All fundamental, dynamical theories accepted so far in physics – Newtonian theory (NT), classical electrodynamics, general relativity, quantum theory, quantum electrodynamics, quantum electroweak theory, quantum chromodynamics and the standard model – can be formulated in terms of a Lagrangian and Hamilton's principle of least action. In the case of NT, this takes the following form. Given any system, we can specify its kinetic energy, KE (energy of motion), and its potential energy, PE (energy of position due to forces), at each instant. This enables us to define the Lagrangian, L, equal at each instant to KE - PE. Hamilton's principle states that, given two instants, $t_1$ and $t_2$, the system evolves in such a way that the sum of instantaneous values of KE - PE, for times between $t_1$ and $t_2$, is a minimum value (or, more accurately, a stationary value, so that it is unaffected to first order by infinitesimal variations in the way the system evolves). From the Lagrangian for NT (a function of the positions and momenta of particles) and Hamilton's principle of least action, we can derive NT in the form familiar from elementary textbooks.

It is this way of formulating NT, in terms of a Lagrangian, L, and Hamilton's principle, that can be generalized to apply to all accepted fundamental theories in physics. Lagrangianism, then, asserts that the universe is such that all phenomena evolve in accordance with Hamilton's principle of least action, formulated in terms of some unified Lagrangian (or Lagrangian density), L. We require, here, that L is not the sum of two or more distinct Lagrangians, with distinct physical interpretations and symmetries, for example one for the electroweak force, one for the strong force, and one for gravitation, as at present; L must have a single physical interpretation, and its symmetries must have an appropriate group structure. We require, in addition, that current quantum field theories and general relativity emerge when appropriate limits are taken.[53]

There are, however, developments in quantum gravity, having to do with "duality", which suggest that Lagrangianism may well be false: see Isham (1997, pp. 194-195). If physicalism(8,1) is true, and space-time has some kind of discrete, quantum mechanical character, then Lagrangianism is false, granted that it requires space-time to be continuous.

An important task for theoretical physics and cosmology would seem to be to formulate the best possible level 3 thesis that is compatible with physicalism(8,1) – compatible, that is with the idea that matter and space-time are unified, space-time, perhaps, having some kind of discontinuous, quantum mechanical character.

(vi) **Cosmological Physicalism**. A basic motivation for making explicit metaphysical assumptions that are implicit in the methods of physics is that it provokes us into inventing new metaphysical possibilities, which we might not otherwise have considered. We are much more victims of *implicit* assumptions – of assumptions we deny making – than of assumptions we make explicit. This consideration prompts the question: Are alternatives to physicalism(8,1) conceivable? There are, of course, endlessly many possible universes less comprehensible physically than those depicted by physicalism(8,1).[54] Are there any different from physicalism(8,1), but just as comprehensible physically?

Physicalism(8,1) holds that the cosmos, at any given instant,[55] is made up of two distinct aspects, which we may call **U** and **V**. **U** is what is depicted by the true physical theory of everything, T. It is inherent in all phenomena, everywhere, at all times. It does not itself change, but determines (perhaps probabilistically) the way that which changes does change. **V**, by contrast, is what does change and vary, from moment to moment, and from one place to the next. **U** and **V** together, at one instant, determine (perhaps probabilistically) **V** at the next instant.

This distinction between **U** and **V** can be traced back to atomism, the very first version of physicalism put forward by Democritus some two and a half thousand years ago. Given atomism, **U** consists of the unchanging properties of atoms and space, while **V** consists of the changing (relative) positions and motions of the atoms. As modern physics developed, ideas about the nature of **U** and **V** have changed, but the distinction itself has persisted up to the present. After Newton, rigid atoms interacting only by contact were transformed into point-atoms surrounded by rigid, centrally-directed fields of force. Here, **U** consists of the unchanging properties of the point-atoms and their surrounding fields of force (including the way the force falls off with distance and the affect it has on other point-atoms), while **V** consists of the changing (relative) positions and motions of the point-particles. Then it emerged, as a result of Faraday's speculations, Maxwell's theory of the electromagnetic field and Einstein's theory of special relativity, that force fields are not rigid. Changes in the field take time to travel. This led to a new *unified field* version of physicalism, according to which everything is made up of an extended, self-interacting, unified field (matter being simply especially intense regions of the field). On the one hand there are changing, variable features of the field, **V**; and on the other, there are the unchanging features of the field, **U**, which determine how **V** changes, and which correspond to the laws of the true theory of the field. Subsequent developments have led to further changes in ideas as to what **U** and **V** are, but have not undermined the distinction itself.

It is no accident that the atomism of Democritus sharply distinguishes **U** and **V**. Atomism arose as an attempt to solve the problem of change, in particular the problem Parmenides posed with his argument that change involves a contradiction, and his view that the cosmos is a homogeneous, unchanging sphere.[56] Parmenides argued that change is impossible because the non-existent cannot exist, hence the world must be full, and hence there can be no room for movement or change. Democritus accepted the argument but rejected the conclusion. There is change, hence the non-existent must exist. The non-existent or, as we might say today, the void surrounds Parmenides's homogeneous, unchanging universe. Other Parmendian universes exist in the void. These can be shrunk down to a minute size, put in relative motion – and we have atomism. Each Democritean atom is a miniature Parmenidean universe. Atomism solves the problem posed by Parmenides by retaining as much as possible of the Parmenidean homogeneous, unchanging universe, but at the same time modifying this view just sufficiently to allow for change and diversity. Atomism solves the general problem of change – the problem of understanding how something can both remain the same *and* change – by segregating very precisely those aspects of atoms which do not change, and those which do, the key to the distinction between **U** and **V**.

But there is another possible response to Parmenides. The cosmos as depicted by Parmenides – a homogeneous unchanging sphere – is a very special, uniquely unified state of the cosmos, the big bang state. This unified, initial state of the cosmos is unstable: spontaneous symmetry breaking occurs, and the cosmos evolves into a state made up of a great number of *virtual* big bang states. Today, every space-time point is made up of just one thing: a fleetingly existent, virtual big bang state.

Quantum theory can be interpreted as asserting that for very short intervals of time there is uncertainty of energy, and this permits so-called *virtual* particles to come into existence in the vacuum and almost immediately cease to exist. According to *cosmic physicalism* – the alternative to atomism as a response to Parmenides – every miniscule space-time region is composed, not of virtual particles, but of the virtual universe in its initial, unified, Parmenidean state. Before the big bang, unity is real and all disunity is virtual. After the big bang, disunity is real and unity is virtual. In a sense, there is only the big bang state. Variety and change come from the different ways in which instantaneously existent, virtual big bang states of the universe are inter-related.

There are, then, two distinct versions of physicalism which we may call *atomistic* and *cosmic* physicalism. They can be regarded as arising as a result of giving different responses to the challenge posed by Parmenides's impossible physically comprehensible universe.

Atomistic physicalism takes the Parmenidean cosmos to depict **U** – that aspect of the cosmos which does not change and which determines the way that which changes, **V**, does change. Initially, **U** represented the properties of the atom. Subsequent developments in theoretical physics have transformed **U**, so that it may be taken to represent the invariant properties of a unified field, a quantum field, space-time of variable curvature, and so on. Despite these developments, the distinction between **U** and **V** persists, and it is this which is the hallmark of atomistic physicalism.

Cosmic physicalism, by contrast, takes the Parmenidean cosmos to be a special, uniquely unified state of the cosmos – the big bang state. According to cosmic physicalism, the true theory of everything, T, specifies the properties of the cosmos in this state. At a fundamental level, the distinction between **U** and **V** does not arise. The distinction only arises when spontaneous symmetry breaking has occurred, and the cosmos consists of momentarily existing virtual big bang states. **V** consists of the different, changing ways in which these momentarily existing big bang states are inter-related.

There are other striking differences between these two versions of physicalism. Cosmic physicalism is inherently *cosmological* in character, whereas atomistic physicalism is not. According to cosmic physicalism, T of itself specifies the initial state of the cosmos, whereas according to atomistic physicalism, initial conditions are required in addition to T to specify the initial state of the cosmos. Cosmic physicalism is inherently *probabilistic*, since spontaneous symmetry breaking is an inherently probabilistic process, whereas atomistic physicalism may be deterministic or probabilistic. Cosmic physicalism must be quantum mechanical to the extent, at least, of incorporating the quantum mechanical distinction between *actual* and *virtual*. Atomistic physicalism makes no such demand.

The two versions of physicalism specify very different conditions for underlying unity to become apparent in as simple a way as possible. According to atomistic physicalism, this happens when the physical system being considered is as simple as possible – the vacuum, or a one particle system or, somewhat more complex, a two particle system. According to cosmic physicalism, it is exactly the opposite: underlying unity is made manifest in a system consisting of *everything* – the entire cosmos in a very special state, the initial big bang state.

Theoretical physics so far has presupposed atomistic physicalism. But it is possible that cosmic, and not atomistic, physicalism is true. Here are a few developments in theoretical physics during the 20$^{th}$ century which may be taken as pointing towards cosmic physicalism.

(a) A basic idea of atomistic physicalism is that the physically simplest, most elemental state of affairs that can exist is the vacuum: physical states of affairs become progressively more complex as 1, or 2, or ... n atoms, or fundamental particles, are added to the vacuum. With the advent of field theory, however, this straightforward ordering of physical complexity begins to break down: here, presumably, we would have to say that the simplest state obtains when the value of the field is everywhere zero, a less simple state arising when the value of the field is everywhere a constant value, more complicated states arising with increasingly complicated variable values of the field.

(b) The idea that the vacuum is the simplest state, in that it is always present in an unchanging form, breaks down further with the advent of Einstein's general theory of relativity (GR). According to GR, the curvature of space-time varies with varying amounts of matter, mass or energy-density; and curved space-time itself possesses energy. Space is no longer a bland, unchanging arena within which more or less complex physical events unfold: the variable curvature of space, or of space-time, itself takes part in dynamical evolution. (Empty, flat, unchanging space-time still arises, however, as a possible solution to the equations of GR.)

(c) With the advent of quantum field theory (QFT), empty space becomes even more complex in that it is full of so-called vacuum fluctuations. These may be pictured as follows. According to one (perhaps somewhat dubious) way of interpreting the time-energy uncertainty relations, $\Delta t \Delta E \geq h/2\pi$ (here, h is Planck's constant), given any state of affairs, for very short time-intervals, $\Delta t$, there will be an uncertainty of energy, $\Delta E$, with $\Delta E \approx h/(2\pi \Delta t)$. Thus, even in empty space, there is sufficient energy, $\Delta E$, available to create an electron/positron pair anywhere, at any time, with the proviso only that such a pair must mutually annihilate after a time $\Delta t \approx h/(2\pi \Delta E)$ and $\Delta E \approx 2m_e c^2$, where $m_e$ is the rest mass of the electron. Such fleetingly existing particles are called "virtual" particles. According to QFT, each minute space-time region is full of virtual processes, involving the creation and annihilation of particle/anti-particle pairs. Indeed, all possible virtual processes occur that violate no other conservation law except that of energy (understood classically). Within very tiny space-time regions so much (virtual) energy can exist that there is sufficient to create a short-lived virtual black hole.

According to QFT, in other words, the vacuum is a mass of seething activity, which averages out to nothing over sufficiently large space-time regions. One may, indeed, interpret QFT as a theory of the vacuum, all possible physical processes going on, within minute space-time regions, as *virtual* processes. In supplying discrete units of energy, $\Delta E_1, \Delta E_2, ... \Delta E_n$, we merely change some of the *virtual* processes into *actual* processes. There is a sense in which the most complex physical state imaginable is no more complex than the vacuum: it is just an energetic state of the vacuum such that some virtual processes are actual processes.[57]

In brief, QFT transforms the simple, elemental, unchanging vacuum of 19th century physics into a seething mass of complex, virtual processes – mirroring, in a ghost-like way, all the complexity of the most complex actual physical processes that exist when there is matter.

It might be supposed that quantum vacuum fluctuations are not real physical phenomena, but only artefacts of the formalism of QFT given a certain (questionable) interpretation. But this does not take into account that vacuum fluctuations have been detected! This was done decades ago by means of the Casimir effect. According to QFT, if two metal plates are held a small distance apart, virtual processes that involve the creation and annihilation of electron/positron pairs will tend to be *suppressed* in the space between the plates. This results in there being a small pressure tending to push the plates together, due to the *unsuppressed* virtual processes taking place in the space surrounding the two plates. This minute force, due to vacuum fluctuations – the Casimir effect – has been detected and measured.

(d) Cosmic physicalism requires that there is a special state of the entire cosmos which is such that all diversity and change disappears: the very distinction between space and matter, we may presume, disappears, there being just *one* homogeneous, instantaneously unchanging *something*, which is also *everything*. This is made possible by big bang cosmology. It is conceivable that the big bang state of the universe, when all of space and matter was packed into a tiny region, was the Parmenidean state of instantaneous unity and homogeneity.

(e) According to GR, the force of gravity is not a force at all; it is rather the tendency of the curvature of space-time to be affected by the presence of matter, or energy-density. With GR, one apparent force, gravity, becomes a feature of physical geometry. This suggests that it may be possible to carry this process of "geometricizing" physics further, the eventual outcome being the unification of space-time, on the one hand, and matter or energy on the other hand. Just this is required by cosmic physicalism.

   (f) According to the Salam-Weinberg theory of the electroweak force (QEWD), at high energies, the distinction between the electromagnetic force on the one hand, and the weak force on the other, disappears. As we go backwards in time towards the big bang, towards a time when the energy-density of the universe was sufficiently high, there existed just one unified force, the electroweak force. As the universe expanded, and the energy-density went down, the unity of the electroweak force was broken: the currently observed disunity of two forces with very different properties emerged. This is strikingly in accord with the basic idea of cosmic physicalism: as we move backwards to the original big bang state so we move towards a state of affairs of greater unity, simplicity, symmetry or homogeneity. The potentially immensely important idea of cosmic spontaneous symmetry breaking, which the Salam-Weinberg theory introduces at the level of fundamental theoretical physics, is precisely what is required to make cosmic physicalism a possibility. It is this development in theoretical physics, above all, which makes cosmic physicalism a viable possibility.

   (g) The idea of an initial state of high symmetry or unity evolving into something asymmetrical and disunified, is further supported by superstring theory – or M theory as it is often called. According to this theory, space has 10 dimensions. The dimensions that we do not observe are curled up into such a minute multi-dimensional "ball" that we do not ordinarily notice their existence. The idea here is that at the big bang state all 10 dimensions of space were curled up in this fashion; the subsequent evolution of the universe consists of just three spatial dimensions growing in size to become, eventually, the space in which we find ourselves. Here, an original Parmenidean unity becomes disunity as three dimensions of space become dramatically different from the rest.

   (vii) **Criteria for Unity**. One of the great merits of aim-oriented empiricism (AOE) is that it provides precise criteria for assessing degrees of unity of dynamical theories of physics. These criteria are incompatible with standard empiricism (SE). Thus, in demanding, for example that, in order to be acceptable, a theory must at least be unified in a type (3) way, with $N = 1$, we thereby commit physics to accepting, as a part of scientific knowledge, the metaphysical thesis that no dynamically unique, spatially restricted entities exist. And similar points arise in connection with the other seven kinds of unity. AOE is required in order to render acceptable the eight criteria for unity spelled out in section 4 above. As long as SE is taken for granted, criteria for unity can only be deployed in a somewhat hypocritical and inconsistent fashion.

(viii) **Rational Discovery**. AOE provides physics with a fallible, non-mechanical but *rational* method for the discovery of revolutionary new physical theories. This involves modifying theories and metaphysical theses, at levels 2, 3 and perhaps 4, in an attempt to resolve conflicts between levels (and between theories at level 2), until it proves possible to formulate a thesis at level 3 with sufficient precision to become a new testable physical theory. It was essentially this rational method of discovery that was employed by Einstein in creating special and general relativity (Maxwell, 1993, pp. 275-305). Nothing like this is at all possible granted SE. From this standpoint, the only scientifically rational way to assess metaphysical ideas would be to favour those ideas that are compatible with existing well-confirmed physical theories. But good new fundamental physical theories, and their associated metaphysical ideas, almost always *contradict* pre-existing theories.  SE leads us to favour precisely the wrong kind of metaphysical idea, from the standpoint of developing a good new scientific theory!  No wonder such stalwart defenders of standard empiricism as Popper (1959, p. 31) and Reichenbach (1938, pp. 381-383) both deny the possibility of a rational method of discovery for science.  No wonder that most physicists and philosophers of physics find the whole process of the discovery of fundamental new physical theories a profound mystery!  (See, for example, Wigner, 1970, ch. 17.)  Only the covert implementation of AOE in scientific practice, under the smokescreen of hypocritical allegiance to SE, has made it possible for some physicists to discover revolutionary new theories.  Repudiation of SE, and the full, explicit implementation of AOE in scientific practice should enable physics to be even more successful than it has been up till now.

(ix) **Pessimistic Induction**.  Theoretical physics tends to advance from one false theory to another.  Thus Galileo's laws of terrestrial motion and Kepler's laws of planetary motion are corrected by Newtonian theory, in turn corrected by general relativity.  Classical physics is corrected by quantum theory, in turn corrected by quantum field theory.  Philosophers of science, viewing this matter from a standard empiricist perspective, tend to regard the fact that physics advances from one false theory to another as having very negative implications for scientific progress.  That physics will continue in this way has even been dubbed "the pessimistic induction" (Newton-Smith, 1981, p. 14).  But viewed from the perspective of AOE, this manner of progression is actually to be expected, if physics really is making progress, and the cosmos really is physically comprehensible.  For, if a theory, $T_o$, is precisely true throughout some restricted domain of phenomena D then, granted physicalism,[58] $T_o$ must specify precisely what does not change, **U**, throughout all phenomena in D, and the way **U** determines how things change in D.  But, according to physicalism, **U** exists unchanged throughout all phenomena.  Thus, if $T_o$ specifies the nature of **U** in D, it will be a straightforward matter to extend $T_o$ so that it specifies **U** for all physically possible phenomena, $T_o$ thus becoming the true theory of everything, T.  Conversely, if $T_o$ cannot be extended in this way to apply correctly to all phenomena, then $T_o$ cannot be precisely true within D: $T_o$ must be false.  In brief, physicalism implies that a dynamical physical theory can only be precisely true of *anything* if it is (capable of being) precisely true of *everything*.

Granted, then, that physics proceeds, not by attaining T in one bound, but rather by developing a succession of theories that apply, with ever increasing accuracy, to ever wider ranges of phenomena until eventually a theory of everything is attained, it is inevitable, granted physicalism, that physics will progress by the development of theories that are all *false* throughout their domains of application until the ultimate, unified true theory of everything is attained (which will be precisely true about everything).[59]  Since physicalism predicts that physics will progress in this way, the fact that physics has so far thus progressed can only count in favour of physicalism: it cannot count against physicalism and AOE, as some have supposed.[60]

(x) **Natural Philosophy**.  Modern science began as natural philosophy – or "experimental philosophy" as it was sometimes called.  In the time of Newton, in the 17$^{th}$ century, science was not only called "natural philosophy".  It was conceived of, and pursued, as a development of philosophy.  It brought together physics, chemistry and other branches of natural science as we know it today, with diverse branches of philosophy: metaphysics, epistemology, methodology, philosophy of science – even theology.  Science and philosophy, which we see today as distinct, in those days interacted with one another and formed the integrated enterprise of natural philosophy.[61]

But then the idea came to prevail – in part because of the impact of misunderstandings about Newton's views about scientific method – that evidence alone determines what is to be accepted and rejected in natural philosophy, or natural *science* as it came to be called.[62]  As a result of the widespread acceptance of this view – standard empiricism as I have called it here – empirical *science* came to be dissociated from non-empirical *philosophy*, and natural philosophy disintegrated (although the name still lingers on in some Scottish universities).

The conception of science responsible for this disintegration of natural philosophy is however, as we have seen, untenable.  Science cannot proceed without making highly problematic metaphysical assumptions concerning the knowability and comprehensibility of the cosmos.  If science is to put scientific rationality fully into practice, and improve knowledge about how to improve knowledge in the light of improving knowledge, then it needs to explore and critically assess metaphysical assumptions and associated methods as an integral part of science itself.  The task of articulating and improving problematic aims and methods of science – the philosophy of science in other words – needs to be pursued as a part of science.  We need to recreate natural philosophy.

(xi) **Philosophy of Science**.  Scientists and philosophers of science recognize that there are a number of baffling unsolved problems about the nature of scientific inquiry.  There is the problem of induction, the problem of how theories can be verified by evidence.  There is the problem of specifying precisely what the methods of science are, especially in view of the fact that methods change over time, and from one scientific discipline to another.  There is the problem of specifying non-empirical criteria theories must satisfy, such as simplicity, unity, explanatory power; and the problem of justifying the use of these criteria in science.  There is the problem of explicating what it can mean to say that physics makes progress, given that it advances from one false theory to another (the problem of verisimilitude).  There is the problem of explaining how the discovery of revolutionary new theories is possible, given that such theories tend to conflict with pre-existing theories, and there are infinitely many possibilities to choose from.  Again and again, physicists have invented new theories which subsequently turn out to yield astonishingly accurate empirical predictions, often predictions of entirely new, unsuspected phenomena.[63]  How is this possible?  There is the mystery of the applicability of mathematics to the physical world.  Mathematicians on occasions develop new mathematical ideas in a context apparently remote from physics which subsequently turn out to be applicable to the physical world in a way no one initially anticipated.[64]  Again, how is this possible?

Long standing attempts to solve these fundamental problems in the philosophy of science within the framework of standard empiricism have all failed.  They are all readily solved, however, once AOE is accepted.[65]  The long standing failure to solve these problems granted SE, and their straightforward resolution granted AOE, amounts to a devastating argument against SE and for AOE.

(xii) **Education**.  The transition from SE to AOE has implications for science education.  Science needs to be taught as natural philosophy.  University courses in physics need to include some discussion of metaphysical and methodological ideas, some exploration of problems of epistemology and philosophy of science.

(xiii) **Fundamental Problem**.  In moving from SE to AOE we are brought face to face with what is, in many ways, a rather grim vision of reality.  Everything is just physics.  Everything is made up of the all-powerful, invariant physical entity, **U**, present everywhere, throughout all possible phenomena, at all times and places, and the variable physical aspect of things, **V**, which change from instant to instant and place to place.  What then becomes of everything that this seems to leave out?  What becomes of the rich tapestry of the experiential world, the colours, sounds, smells and tactile qualities we experience?  What becomes of our inner sensations, our inner states of consciousness?  What becomes of free will, meaning and value, everything that seems precious in life?  How can any of this exist if the cosmos really is physically comprehensible in the way that this has been explicated here?

One great merit of AOE, I suggest, is that it brings us face to face with this fundamental problem.  It is, in my view, our fundamental problem of both thought and life.  It can be formulated like this: How can our human world, imbued with perceptual qualities, consciousness, free will, meaning and value, exist and best flourish, embedded as it is in the physical universe?  Elsewhere, I have argued that we need to interpret physics as being restricted to depicting what may be called the "causally efficacious" aspect of things – that aspect which determines (perhaps probabilistically) how events unfold.  What things look like and feel like, what it is like to *be* a bit of the physical universe: these experiential aspects of things necessarily lie beyond the scope of physics, and any attempt to extend physics to incorporate the experiential would destroy the marvellously explanatory capacity of physics.  Physics omits all mention of the experiential and the value-laden because it can without curtailing its predictive and explanatory task; and it must do this in order to retain the astonishing explanatory power of its theories.[66]

(xiv) **Wisdom-Inquiry**.  As I have argued at length elsewhere, AOE has revolutionary implications for rationality, for academic inquiry, and for the future of humanity.[67]  It is not just science that has problematic aims; many other endeavours, individual, institutional and even global, have problematic goals as well.  In this essay I have argued that the problematic character of the basic aim of physics means that we need to represent this aim in the form of a hierarchy of aims, thus providing physics with a fixed framework of aims and methods (assumptions and methods) within which much more problematic aims and methods can be improved as physics proceeds.  These considerations can be generalized and applied to any human endeavour with problematic aims.  We need a new conception of rationality – aim-oriented rationality (AOR) – a generalization of AOE, which emphasizes that, whatever we are doing, if our basic aim is problematic, we need to represent it in the form of a hierarchy of aims, thus providing ourselves with a framework of unproblematic aims and associated methods within which more specific and problematic aims and associated methods can be improved as we act.  Applied to academic inquiry as a whole, AOR implies – I have argued – that we need a new kind of inquiry – *wisdom-inquiry* – which puts problems of living at the heart of the enterprise, and takes the basic aim to be to seek and promote wisdom, construed to be the capacity realize what is of value in life for oneself and others, wisdom thus including knowledge, technological know-how and understanding, but much else besides.  AOR applied to our social world implies that much that we do – individually, institutionally and globally – needs to be reorganized so that problematic aims may be acknowledged and improved. Above all, AOR needs to be applied to the profoundly problematic endeavour to make progress towards as good a world as possible.

## 10. Conclusion

Granted the view that evidence alone decides what is to be accepted in science, it is clear that science cannot possibly be said to have established that the cosmos is physically comprehensible. But this view of science is untenable. Physics quite properly persistently only accepts theories that are (a) sufficiently successful empirically, and (b) sufficiently *unified* – even though endlessly many empirically more successful, disunified rivals can always be concocted. This means we must construe physics as accepting, as a part of scientific knowledge, a hierarchy of metaphysical theses concerning the knowability and comprehensibility of the cosmos, these theses becoming less and less substantial, and so more and more likely to be true, as we go up the hierarchy, and also more nearly such that their truth is required for science, or the pursuit of knowledge, to be possible at all. One of these theses is that the cosmos is physically comprehensible – that is, such that the true physical theory of everything is unified. It is, in other words, a part of current theoretical knowledge in physics that the cosmos is physically comprehensible.

This *aim-oriented empiricist* view of science has a range of fruitful implications for science itself, and for our understanding of science. It extends the scope of scientific knowledge and understanding about a matter of great significance: the ultimate nature of the cosmos. It provides us with explicit criteria for the assessment of the extent to which physical theories are *unified*. It provides physics with a rational if fallible method for the discovery of revolutionary new physical theories. It transforms science so that it becomes more nearly like natural philosophy of Newton's time – an enterprise that brings together and integrates empirical science, methodology, philosophy of science, metaphysics and epistemology. It facilitates the solution to a number of fundamental problems in the philosophy of science: problems of induction, verisimilitude, simplicity, and scientific discovery. This view of science helps to highlight the nature of the fundamental problem we face in thought and in life: How can our human world, imbued with all that we hold to be of value, exist and best flourish, embedded as it is in the physical cosmos? Finally, this view of science has implications for views about the nature of rationality, and the nature of rational inquiry quite generally.

**Notes**

[1] Throughout this essay, "metaphysical" means "factual thesis – thesis about the world – that is not empirically testable".

[2] See Maxwell (1972a; 1974; 1984, chs. 5 and 9; 1993; 1998; 2000a; 2002; 2004a; 2005; 2006a; 2007, chs. 5, 9 and especially 14; 2008; 2011a; 2012a; 2012b; 2013).

[3] See Miller (2006, p. 92-94) and Muller (2008).

[4] See Vicente (2010).

[5] For my replies to Miller, Muller and Vicente see Maxwell (2006b; 2009a; 2010a, pp. 667-677).

[6] See for example Kneller (1978, pp. 80-87 & 90-91), Harris (1980, pp. 25-26), Chakravartty (1999), Smart (2000), Juhl (2000), McHenry (2000). Weinert (2000), Roush (2001), Muller (2004), McNiven (2005), Iredale (2005), Grebowicz (2006).

[7] For discussion of the claim that Kuhn and Lakatos defend versions of SE see Maxwell (1998), p. 40, and Maxwell (2005). Bayesianism might seem to reject SE, in acknowledging both prior and posteriori probabilities. But Bayesianism tries to conform to the spirit of SE as much as possible, by regarding prior probabilities as personal, subjective and non-rational, their role in theory choice being reduced as rapidly as possible by empirical testing: see Maxwell (1998), p. 44.

[8] See Feyerabend (1975); Bloor (1991); Barnes, Bloor and Henry (1996).

[9] For more detailed discussion of the point that SE is widely taken for granted see Maxwell (1984), chs. 2 and 6; Maxwell (1998), ch. 2; Maxwell (2004a), pp. 13-14, note 14.

[10] For a more detailed exposition of SE, see Maxwell (2007), pp. 32-51. For grounds for holding scientists do, by and large, accept SE, see Maxwell (1998, pp. 38-45; 2007, pp. 145-156; 2004a, pp. 5-6, note 5).

[11] The famous arguments of David Hume (1959) concerning causation have long rendered familiar the point that, for all we can ever know for certain, physical laws may, at some future time, abruptly change – or, to put the same point in an equivalent way, given any physical theory, however empirically successful, there will always be infinitely many rival theories that fit all available evidence just as well but which postulate an abrupt change in physical laws at some future time. What I go on to argue in the text is

that these "Humean" theories, as we may call them, that postulate an abrupt change in physical law at some future time are just special cases of a much wider class of disunified theories that postulate an abrupt change in physical law as conditions change in some way other than time, such as kind of system, or experimental conditions. And furthermore, I argue, all these disunified theories can be further doctored to be empirically *more successful* than the theories we accept.

[12] Even if no phenomena ostensibly refute T, each of $T_1^*, T_2^*, \ldots T_\infty^*$ is still more successful empirically than T.

[13] For earlier refutations of SE along these lines, see Maxwell (1998, ch. 2; 2004a, ch.1; 2005; 2007, ch. 9). For my rebuttal of sixteen objections to the validity of this refutation of SE see Maxwell (2012c).

[14] Richard Feynman has provided the following amusing illustration of this point (Feynman et al. 1965, 25-10 – 25-11). Consider an appallingly disunified, complex theory, made up of $10^{10}$ quite different, distinct laws, stuck arbitrarily together. Such a theory can easily be reformulated so that it reduces to the dazzlingly unified, simple form: $A = 0$. Suppose the $10^{10}$ distinct laws of the universe are: (1) $F = ma$; (2) $F = Gm_1m_2/d^2$; and so on, for all $10^{10}$ laws. Let $A_1 = (F - ma)^2$, $A_2 = (F - Gm_1m_2/d^2)^2$, and so on. Let $A = A_1 + A_2 + \ldots + A_{10^{10}}$. The theory can now be formulated in the unified, simple form $A = 0$. (This is true if and only if each $A_r = 0$, for $r = 1, 2, \ldots 10^{10}$).

[15] See for example Salmon (1989). Elsewhere, I have formulated seven problems of unity or simplicity: see Maxwell (1998, pp. 104-105).

[16] For my criticisms of proposals put forward by Jeffreys and Wrinch (1921), Popper (1959, pp. 62-70 & 126-145), Friedman (1974), Kitcher (1981) and Watkins (1984, pp. 203-213), see my (1998, pp. 56-68). For criticisms of more recent proposals see my (2004b).

[17] For earlier accounts of my proposed solution to the problem of unity of physical theory see Maxwell (1998, chs. 3 and 4; 2004a, chs. 1-2 and appendix; 2004b; 2007, pp. 373-386; and 2011a).

[18] As I have formulated it here, (1) is open to two somewhat different interpretations. First, for N = 1 we require only that *the same* law operates throughout space in the sense that this would be true even if the law in question asserted that all objects experience a force directed at a unique point in space, and inversely proportional to their distance from that point. Second, for N = 1, we require that *the same* law operates throughout space in the sense that a mere change of position in space of an isolated physical system has no effect on the way the system evolves. An analogous distinction arises in connection with time. In what follows I adopt the second interpretation of (1), which means that a theory which is unified with respect to (1) exhibits symmetry with respect to spatial location, and time of occurrence. As far as the *ad hoc* version of NT is concerned, N = 2 for both versions of (1).

[19] Counting entities is rendered a little less ambiguous if a system of M particles is counted as a (somewhat peculiar) field. This means that M particles all of the same kind (i.e. with the same dynamic properties) is counted as *one* entity. In the text I continue to adopt the convention that M particles all the same dynamically represent one *kind* of entity, rather than one entity.

[20] An informal sketch of these matters is given in (Maxwell, 1998, ch. 4, sections 11 to

[13], and the appendix).  For rather more detailed accounts of the locally gauge invariant structure of quantum field theories see: (Moriyasu, 1983; Aitchison and Hey, 1982, part III; and Griffiths, 1987, ch. 11).  For a non-technical discussion of the role of symmetry and group theory in physics, see (Maxwell, 1998, pp. 257-265); for somewhat more technical introductory accounts of group theory as it arises in physics see (Isham, 1989; or Jones, 1990).

[21] For accounts of spontaneous symmetry breaking see (Moriyasu, 1983; Mandl and Shaw, 1984; Griffiths, 1987, ch. 11).

[22] Some kinds of disunity can be decreased in a trivial way by splitting a theory into two or more distinct theories.  Thus it may be possible to split a theory postulating two distinct forces into two distinct theories, each postulating one force.  This trivial manoeuvre can be outlawed by insisting that unity applies to all fundamental physical theories applying to the full range of physical phenomena – and to empirical laws themselves if some phenomena are without a theory that applies to them.  A new theory is then acceptable, on unity grounds, if its acceptance improves the unity of the totality of fundamental theory.

[23] The account of theoretical unity given here simplifies the account given in (Maxwell, 1998, chs. 3 and 4), where unity is explicated as "exemplifying physicalism", where physicalism is the metaphysical thesis asserting that the universe has some kind of unified dynamic structure.  Explicating unity in that way invites the charge of circularity, a charge that is not actually valid (see Maxwell, 1998, pp. 118-23 and 168-72).  The account given here forestalls this charge from the outset.

[24] This point is of fundamental importance for the problem of induction.  Traditionally, the problem is interpreted as the problem of justifying exclusion of empirically successful theories that are *ad hoc* in sense (1): How can evidence from the past provide grounds for any belief about the future?  This makes the problem seem highly "philosophical", remote from any problem realistically encountered in scientific practice.  But the moment it is appreciated that the problem of justifying exclusion of empirically successful theories that are *ad hoc* in sense (1) is just an extreme, special case of the more general problem of excluding empirically successful theories that are *ad hoc* in senses (1) to (8), it becomes clear that this latter problem is a scientific problem, a problem of theoretical physics itself.  For the implications of this crucial insight, and for a proposal as to how the problem of induction is to be solved exploiting it, see Maxwell (1998, especially chs. 4 and 5; 2004a, pp. 205-220; 2007, pp. 400-430).  See also Maxwell (2006a).

[25] For further details see Maxwell, 1998, pp. 110-113.

[26] For a fascinating discussion of the problems that arise in connection with the wider notion of what I have called 'invariant function', see Roger Penrose's discussion of what he calls the 'Eulerian' notion of function: Penrose (2004, 6.4).

[27] See Boscovich (1966).

[28] Strictly speaking, Newton's theory clashes with Boscovich's metaphysics, because Newton's theory implies that the force between point-particles becomes infinitely attractive as the distance between them becomes zero.  Boscovich's view can easily be reformulated to remove this clash.  Point-particles can be assigned different charges, electrical, and perhaps other charges in addition to gravitational charge.

[29] It must be admitted, however, that the field introduces new problems, especially if

charged point-particles are embedded in the field. Problems of infinite self-interaction arise, problems that haunt, not just classical electrodynamics, but quantum electrodynamics, and other quantum field theories as well.

[30] I first expounded and defended a version of this hierarchical view in Maxwell (1974). It was further elaborated in Maxwell (1984 and 1993). A more elaborate version still is expounded and defended in great detail in Maxwell (1998). For a more detailed defence of the version indicated here, see Maxwell (2004a, chs. 1 and 2, and appendix). In Maxwell (2005) I argue that this view is a sort of synthesis of the views of Popper, Kuhn and Lakatos, but an improvement over the views of all three. In Maxwell (2007, ch. 14) I give a detailed exposition of AOE, and argue, in some detail, that AOE succeeds in solving major problems in the philosophy of science, including the problems of induction, simplicity and verisimilitude. For an account of how I developed AOE, partly as a result of criticizing Popper's falsificationism, see Maxwell (2012a).

[31] See Maxwell (1993, pp. 275-305).

[32] See Einstein (1973, p. 357). See, also, Maxwell (1993, pp. 275-305).

[33] AOE can be generalized in a number of ways. First, it can be reformulated so that it depicts a hierarchy of *aims* and methods, rather than *assumptions* and methods. We can interpret the aim of physics, at each level, from 3 to 7, to be to turn the assumption, at that level, into the precise, true physical theory of everything. Second, we need to recognise that there are, not just *metaphysical* assumptions inherent in the aims of science, but also *value* assumptions, and *humanitarian* or *political* assumptions. The level 4 aim of seeking explanatory truth – truth presupposed to be explanatory – is a special case of the more general aim of seeking truth that is *of value*, intellectually or practically. And this in turn is sought so that it may be *used* by people to enrich life, in either cultural or practical ways. These value and political assumptions inherent in the aims of science are almost more problematic than metaphysical assumptions. Here, too, aims and assumptions need to be made explicit so that they may be critically assessed and, we may hope, improved. Third, AOE can be generalized so that it applies to *all* the diverse branches of natural science, and not just to theoretical physics. Each science has its own specific, problematic aims which need to be made explicit and subjected to sustained critical scrutiny. Finally, it is not just in science that aims are problematic; this is the case in life too, at all levels, from the individual and personal, to the institutional, social and global. AOE needs to be generalized to become a conception of rationality – *aim-oriented rationality* – which stresses that problematic aims need to be represented in the form of a hierarchy, so that aims can be improved as we act, as we live. For these generalizations of AOE, and their implications for academic thought and life, see: Maxwell (1976a; 1984 or 2007; 2004a; 20010a; 2012a).

[34] Popper (1959; 1963).

[35] Kant argued that we can know nothing about the noumenal world except that it exists. But his argument is not just invalid; his position is inconsistent. To assert that the noumenal world is such that we can know nothing about it is to assert that we do know something about it, namely that it is such that it is unknowable. Kant was presupposing a version of Cartesian dualism, much discredited since Kant's time, according to which appearances form an impenetrable barrier between us and reality, preventing us from acquiring any knowledge of reality. We cannot possibly *know* that we cannot know anything about the nature of reality (unless we interpret "know" in an illegitimately epistemologically strong way to mean something like "know with absolute certainty").

How, we may wonder, could Kant possibly be so certain that reality is such that it is unknowable? We need to be more epistemologically modest, and admit that reality may, for all we know, be such that it is knowable. Here is an item of knowledge about ultimate reality, Kant's noumenal world, which we can be reasonably confident is true: *the cosmos is not a chicken*. Just possibly this is false, and the milky way and the other galaxies we observe are giant molecules in the wattle of the cosmic chicken. Almost certainly, this is not the case. We can be reasonably certain of the truth of *the cosmos is not a chicken* because the statement says so little. There are infinitely many ways in which the cosmos can contrive not to be a chicken.

[36] Remark Einstein made during his first visit to Princeton University in 1921.

[37] Some Popperians argue that any attempt to argue for the acceptance of a thesis or theory goes against Popper's falsificationist, critical rationalist philosophy. But this is wrong. An argument intended to justify *acceptance* of a thesis sets out to do something quite different from an argument intended to justify the *truth* (or probable truth) of a thesis. Popper himself argues for the acceptance of theory that has greater empirical content than its predecessor, and has survived severe testing – a central claim of Popper (1959). Here, the more acceptable a theory is, according to Popper, so the less likely it is to be true – just because its empirical content is greater. One could hardly highlight more dramatically the distinction between arguing for *acceptance* of a thesis, and arguing for its *truth*.

[38] What might we have to lose? If meta-knowability is false, then efforts to improve knowledge about how to improve knowledge cannot succeed, and will have turned out to have been for nothing. On the other hand, the only way we could discover that meta-knowability is false is to try to improve knowledge about how to improve knowledge, and fail in the attempt.

[39] No circularity is involved here, because no attempt is made to justify acceptance of meta-knowability by an appeal to the success of science. The argument is that we should make meta-knowability explicit even if GAOE science is entirely unsuccessful. We should make explicit metaphysical assumptions implicit in our methodological practice whether this practice has met with real success or only apparent success.

[40] But there is another argument, as we have seen. In accepting meta-knowability, we have little to lose and may have much to gain.

[41] Comprehensibility may *promise* meta-knowability, but it does not *imply* it. The cosmos might be comprehensible but unknowable to us; or it might be comprehensible in such a way that it is not possible for us to put GAOE into practice in an empirically successful way.

[42] For further discussion see Maxwell (1998, pp. 80-89, 131-140, 257-265 and additional works referred to therein).

[43] The claim that science has "established" that the universe is physically comprehensible is perhaps a bit misleading, in that it strongly suggests that science has established the (probable) truth of the thesis. What I have done, put more accurately, is to show that the thesis is a central part of current theoretical scientific knowledge. The thesis is so firmly accepted that accepted physical theories that clash with it, such as the standard model and general relativity, are regarded in science as problematic, and likely to turn out to be false. Theories that clash violently with the thesis, in that they are

seriously disunified, are not even considered in physics even though they are ostensibly empirically more successful than accepted unified rival theories.

[44.] For my views as to how our human world, imbued with colours, sounds, meaning and value, consciousness and free will can be accommodated within the universe, presumed to be physically comprehensible, see Maxwell (1966; 1968a; 1968b; 1976a; 1984 or 2007, ch. 10; 2000b; 2001; 2009b; 2011b; and especially 2010a).

[45] This isn't quite correct. Relativistic phenomena refute non-relativistic OQT. These phenomena can, however, at least in principle, be predicted by relativistic quantum theory.

[46] See, for example, Bohr (1949).

[47] See Maxwell (1972b; 1976b, pp. 276-279; 1988, pp. 1-8; 1998, pp. 228-235).

[48] See Maxwell (1993, pp. 290-296).

[49] See Maxwell (1988, pp. 1-8; 1998, pp. 230-235).

[50] See Maxwell (1976b; 1982; 1988; 1994; 1998; 2004; 2011c).

[51] Maxwell (1994; 1998, ch. 7; 2004c; 2011c).

[52] I know of only three theoretical physicists who have sought to develop fully micro-realistic, fundamentally probabilistic versions of quantum theory. The first is Roger Penrose: see his (1986; 2004, ch. 30). The other two are Gordon Fleming and Rudolf Haag (personal communication).

[53.] For accounts of Lagrangian formulations of classical and quantum mechanical theories see: Feynman et al. (1963: vol. II, ch. 19), Mandl and Shaw (1984: ch. 2), Goldstein (1980).

[54] Universes that just meet the requirements of physicalism(n,1) become less and less comprehensible physically as n = 7, 6 … 2,1. And, for any given n, physicalism(n,N) with N = 1, 2,… represent universes increasingly disunified, and hence increasingly incomprehensible physically.

[55] Or on any given spacelike hypersurface, looking at things from the standpoint of general relativity.

[56] This story has been told brilliantly by Karl Popper: see Popper (1998).

[57.] For an excellent account of the quantum theory of the vacuum, see Aitchison (1985).

[58] 'Physicalism' here means 'physicalism(n,1) with $8 \geq n \geq 5$'. We require a version of physicalism which asserts that there is an invariant **U** throughout all phenomena that are physically possible (according to that version of physicalism).

[59] Or rather, precisely true about that aspect of what exists which determines the way events evolve everywhere, at all times, throughout all phenomena.

[60] See, for example, Laudan (1980), Newton-Smith (1981).

[61] This point was well made long ago by Burtt, (1932).

[62] Newton came to be interpreted as advocating that evidence alone should determine what theories are to be accepted – even that theories should be arrived at by inductive reasoning from phenomena. But Newton formulated three of his four rules of reasoning in such a way that it is clear that these rules make assumptions about the nature of the universe. Thus rule 1 asserts: "*We are to admit no more causes of natural things than such as are both true and sufficient to explain their appearances.*" And Newton adds: "To this purpose the philosophers say that nature does nothing in vain, and more is in vain when less will serve; for Nature is pleased with simplicity, and affects not the pomp of superfluous causes": see Newton (1962, vol. 2, p. 398). Newton understood that

persistently preferring simple theories means that Nature herself is being persistently assumed to be simple (which violates standard empiricism).  In so far as Newton's views about method played a major role in the general acceptance of standard empiricism after his death, it has to be said that it was a radical misunderstanding of these views that played this role.  Newton's actual views on method are more sophisticated than all those 20$^{th}$ and 21$^{st}$ century scientists and philosophers of science who have taken one or other version of standard empiricism for granted.

[63] Especially notable in this respect are: Newtonian theory, Maxwellian electrodynamics, Einstein's special and general theories of relativity, Bohr's quantum theory of the atom, Heisenberg's and Schrödinger's non-relativistic quantum theory, Dirac's relativistic quantum theory of the electron, quantum electrodynamics of Tomonaga, Schwinger, Feynman and Dyson, and quantum electroweak dynamics and quantum chromodynamics developed by Yang, Mills, Salam, Weinberg, Gell-Mann and others.

[64] Especially noteworthy is Appollonius's characterization of conic sections over one and a half thousand years before Galileo and Kepler employed conic sections (the parabola and the ellipse) to describe terrestrial and planetary motion.  Noteworthy, too, is Riemann's invention of Riemannian geometry some 60 years before Einstein depicted space-time as Riemannian in his general theory of relativity.

[65] How AOE solves the problem of induction is outlined above, in sections 7 and 8.  There, I justified acceptance of theses at levels 8 to 4 as a part of theoretical scientific knowledge.  Our current best fundamental theories in physics – general relativity and the standard model – are to be accepted because (a) they have met with sufficient empirical success, and (b) they accord sufficiently well with the thesis at level 4 in the hierarchy of theses of AOE.  For earlier, more detailed accounts of how AOE solves the problem of induction, in each case adding to and improving earlier accounts, see Maxwell (2007, pp. 400-430; 2004a, pp. 205-220; 1998, ch. 5).  See also Maxwell (2006a).  It might be thought that AOE only specifies the nature of scientific method as far as theoretical physics is concerned.  But, as I indicted in note 33, AOE can readily be generalized to apply to all branches of natural science, and to natural science as a whole: see Maxwell (2004a, ch. 2, especially pp. 41-47).  Different sciences have somewhat different aims; and aims evolve as scientific knowledge evolves.  Because of its meta-methodological character, AOE can do justice to the diversity of aims and methods, as one goes from one science to another, and from one time to another within one science; at the same time, AOE can do justice to the unity of method throughout this diversity of methods of natural science.  How criteria of simplicity, unity and explanatory power can be explicated within the framework of AOE has been indicated in section 4 above.  See also Maxwell (1998, chs. 3 and 4; 2004a, pp. 160-174).  For my proposal as to how one can solve the problem of verisimilitude see Maxwell (2007, pp. 393-400 and 430-433).  For my proposed solution to the mystery of the discovery of revolutionary new theories in physics, and the applicability of mathematics to the physical world, see Maxwell (1998, pp. 219-223).  AOE does of course introduce a new problem: Why should the cosmos be physically comprehensible to us?  "The eternal mystery of the universe is its comprehensibility", as Einstein (1973, p. 292) once put it.  For my suggestions as to how this problem can be solved see Maxwell (2001, pp. 254-258).

[66] See works referred to in note 44.
[67] See Maxwell (2007). See also Maxwell (1976a; 2010a).